\documentclass[a4paper,11pt]{article}
\pdfoutput=1 \usepackage{jheppub_2} 
\usepackage[T1]{fontenc} 
\usepackage[utf8]{inputenc}
\usepackage{array}
\usepackage{multirow}
\usepackage{booktabs}
\usepackage{mathrsfs}
\usepackage[euler]{textgreek}

\usepackage{amsmath}
\usepackage{amsthm}
\usepackage{amsfonts}
\usepackage{amssymb}
\usepackage{graphicx}
\usepackage[font={small}]{caption}
\usepackage{float}
\usepackage[export]{adjustbox}
\usepackage[hang, flushmargin,bottom]{footmisc} 
\usepackage{hyperref}
\hypersetup{
     colorlinks   = true,
     citecolor    = blue
}
\usepackage{placeins}
\usepackage{enumitem}
\usepackage{slashed}
\usepackage{bbm}
\usepackage{appendix}

\usepackage{titlesec}
\titleformat{\chapter}[display]
  {\normalfont\LARGE\bfseries}
  {\chaptertitlename\ \thechapter}{5pt}{\LARGE}
  \titlespacing*{\chapter}{0pt}{-20pt}{35pt}
\usepackage{bigstrut}
\usepackage{pst-node}
\usepackage{pstricks}
\usepackage{datetime}
\usepackage{color, colortbl}
\definecolor{GrayLight}{gray}{0.9}

\usepackage{physics}
\usepackage{mathtools}
\usepackage{setspace}
\usepackage{fancyhdr}
\usepackage[makeroom]{cancel}
\newcommand{\be}{\begin{equation}}
\newcommand{\ee}{\end{equation}}
\newcommand{\bes}{\begin{equation*}}
\newcommand{\ees}{\end{equation*}}

\renewcommand\thesection{\arabic{section}}
\usepackage{hyperref}
\hypersetup{%
  colorlinks = true,
  linkcolor  = blue
}
\usepackage{soul}

\usepackage{placeins}
\usepackage{hyperref}
\usepackage{makeidx}
\usepackage{mathrsfs}
\usepackage{dsfont}
\usepackage{fancybox}
\usepackage{amsfonts}
\usepackage{graphicx}
\usepackage{tcolorbox}
\usepackage{marvosym} 
\usepackage{tikz}
\usepackage{tabulary}
\usepackage{pgfplots}
\usepackage{subfig}
\usepackage{amsmath}
\usepackage{slashed}
\usepackage{mathtools}
\usepackage{fancyhdr} 
\usepackage{nccmath}
\usepackage{verbatim}
\usepackage{color}
\usepackage{amsmath, amsthm, amssymb, amsfonts}
\usepackage{array}
\newcolumntype{P}[1]{>{\centering\arraybackslash}p{#1}}
\usepackage{enumerate}
\usepackage{physics}
\usepackage[font={small, up}]{caption}
\usepackage{tikz}
\usetikzlibrary{"arrows", "automata", "backgrounds", "calendar", "chains", "matrix", "mindmap", "patterns", "petri", "shadows", "shapes.geometric", "shapes.misc", "spy", "trees"}
\usepackage{appendix}

\newcommand{\imineq}[2]{\vcenter{\hbox{\includegraphics[height=#2ex]{#1}}}}

\newcommand{\myComment}[1]{}

\newcommand{\beq}{\begin{equation}}
\newcommand{\eeq}{\end{equation}}

\newcommand{\SU}{\,{\rm SU}}

\newcommand{\U}{\,{\rm U}}

\notoc 

\title{\boldmath Baryonic Higgs and Dark Matter}

\author[a]{Pavel Fileviez P\'erez,}
\author[b]{Clara Murgui}
\author[a]{and Alexis D. Plascencia}
\affiliation[a]{Physics Department and Center for Education and Research in Cosmology and Astrophysics (CERCA), 
Case Western Reserve University, Cleveland, OH 44106, USA}
\affiliation[b]{Walter Burke Institute for Theoretical Physics, California Institute of Technology, Pasadena, CA 91125}

\emailAdd{pxf112@case.edu}
\emailAdd{cmurgui@caltech.edu}
\emailAdd{alexis.plascencia@case.edu}

\abstract{We discuss the correlation between dark matter and Higgs decays in gauge theories where the dark matter is predicted from anomaly cancellation. 
In these theories, the Higgs responsible for the breaking of the gauge symmetry generates the mass for the dark matter candidate.
We investigate the Higgs decays in the minimal gauge theory for Baryon number. After imposing the dark matter density and direct detection constraints, we find that the 
new Higgs can have a large branching ratio into two photons or into dark matter. 
Furthermore, we discuss the production channels and the unique signatures at the Large Hadron Collider.}

\begin{document} 

\maketitle


\flushbottom

\newpage 

\newpage

\section{Introduction}
The discovery of the Standard Model (SM) Higgs boson with a mass of 125 GeV at the Large Hadron Collider (LHC) has opened a possible new portal to dark matter~\cite{Silveira:1985rk,McDonald:1993ex,Patt:2006fw}. If dark matter acquires mass through spontaneous symmetry breaking, then it can be expected that dark matter will be coupled to the SM Higgs. The ATLAS and CMS collaborations have throughly search for dark matter by looking for the invisible decays of the SM Higgs boson. The most recent upper bound reported by the ATLAS collaboration from their Run II analysis of data with $\sqrt{s}=13$ TeV and $139 \text{ fb}^{-1}$ luminosity is given by
\begin{equation}
\text{Br} ( h \to \text{invisible} ) < 0.13 \text{ at }95\% \text{ CL}\text{~\cite{ATLAS-CONF-2020-008}.}
\end{equation} 
Moreover, a statistical combination of the results from both Run I and II by ATLAS gives $\text{Br} (h \to \text{invisible}) < 0.11 \, (95\% \text{ CL})$~\cite{ATLAS:2020kdi}, while a combined analysis by the CMS collaboration finds that $\text{Br} ( h \to \text{invisible} ) < 0.19 \, (95 \% \text{ CL})$~\cite{Sirunyan:2018owy}. 

Nevertheless, the discovery of the SM Higgs also motivates the question of whether there are other fundamental scalars to be discovered at the LHC. This bring us to question the origin of the dark matter mass. Among the different possibilities that could give rise to the mass of the dark matter candidate, the idea of generating the dark matter mass through the spontaneous breaking of a given gauge symmetry is specially attractive from a theoretical point of view. In that context, the dark matter mass would be connected to the energy scale of the force acting on it. 

In theories where an anomalous symmetry is promoted to a local symmetry, a dark matter candidate can be predicted from the cancellation of gauge anomalies; for example when gauging baryon or 
lepton number~\cite{Perez:2014qfa,FileviezPerez:2011pt,Duerr:2013dza}. In this case, the dark matter $\chi$  will interact through a new Abelian gauge symmetry $\U(1)_X$  with charge $n_X$. Then, through the following Yukawa interaction 
\begin{equation}
-{\cal L} \supset y_\chi  \chi \chi h_X + \text{h.c.},
\end{equation}
the dark matter can get a mass once the new Higgs $h_X$ acquires a vacuum expectation value (vev) that spontaneously breaks the $\U(1)_X$ and gives mass to the corresponding gauge mediator that determines the scale $\Lambda_X \sim M_{Z_X}$ where the new symmetry is broken. Hence, the dark matter mass is linked to such scale as follows:
\begin{equation}
M_\chi= \frac{\sqrt{2} \, y_\chi}{n_X g_X} M_{Z_X}.
\end{equation}
\begin{figure}[h!]
\centering
\includegraphics[width=0.6\linewidth]{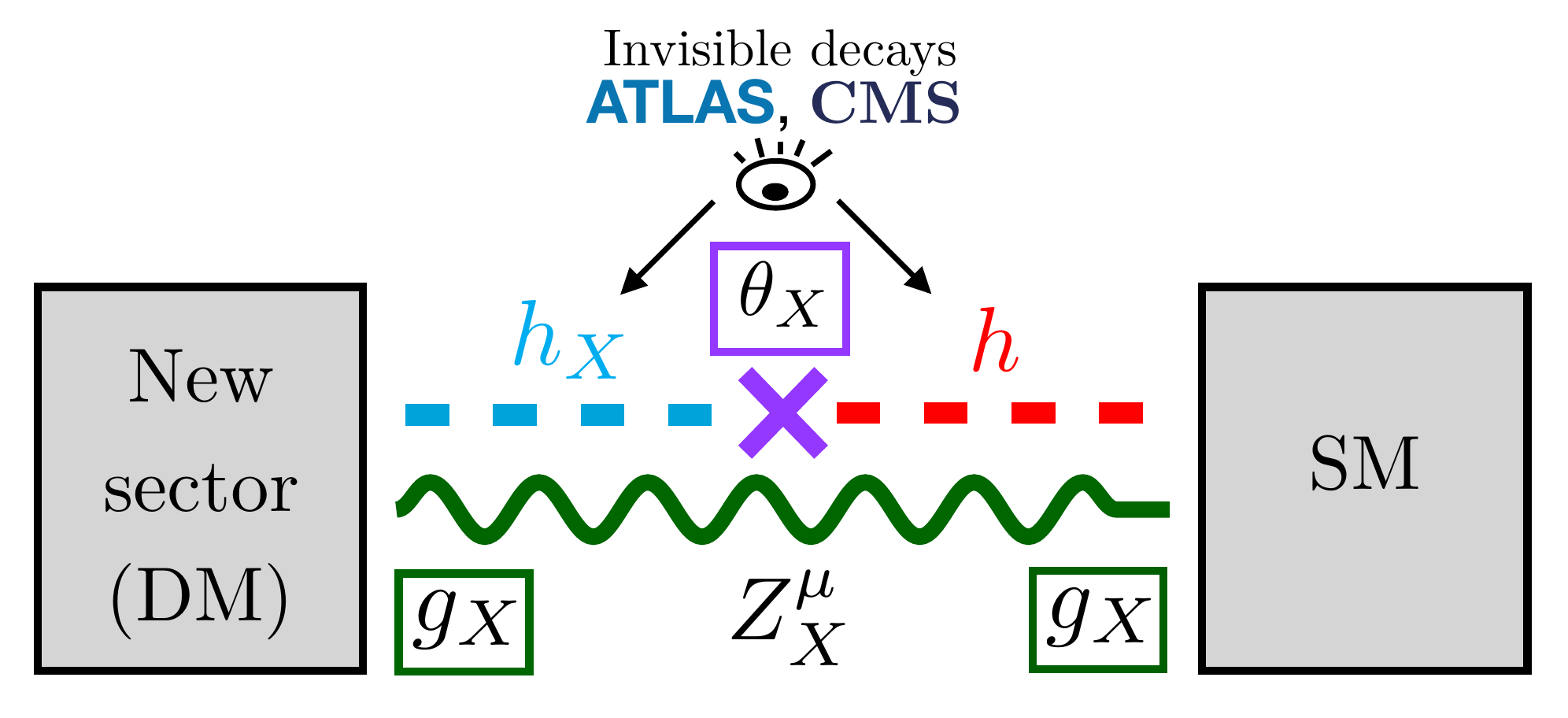} 
\caption{Collider experiments can search for dark matter through the SM Higgs boson decays to invisible. However, in theories where the dynamical origin of the dark matter mass can be understood, those decays can be strongly suppressed by the scalar mixing angle. Consequently, it might be more convenient to look for signals arising from the invisible decays of a the new Higgs, $h_X$, responsible for the dark matter mass and the breaking of the new symmetry $\U(1)_X$.}
\label{fig:connections}
\end{figure}
\indent As we will discuss in the next section, it is possible to define simple anomaly-free theories where the dark matter candidate is predicted by the condition of anomaly cancellation. 
In this context, one would expect that the new Higgs $h_X$ mixes with the SM Higgs through the scalar potential as illustrated in Fig.~\ref{fig:connections}. This mixing can be parametrized by an angle $\theta_X$, which is already constrained by collider experiments to be at most $\sin \theta_X< 0.3$~\cite{Ilnicka:2018def,Adhikari:2020vqo}. However, as we will learn along this manuscript, direct detection bounds on dark matter can place a more stringent bound on the mixing angle between the two scalar bosons.

A dark matter scenario such as the one just described could be seen at the collider searches for invisible decays of the SM Higgs boson, but such signals will always suffer from the mixing suppression of $\sin^2 \theta_X$. Then, the mixing portal might not be the best option to detect dark matter. In this paper we suggest to look for dark matter in alternative portals that will of course depend on the new force we are considering. In the theories we consider, where the dark matter candidate gets mass through a new Higgs mechanism, we will propose ways to study the correlations between the physics of the new Higgs at colliders and the dark matter phenomenology.

In this work we study the motivated scenario in which the new gauge symmetry is identified to be baryon number $\U(1)_B$~\cite{Perez:2014qfa,FileviezPerez:2011pt,Duerr:2013dza}. Our study focuses on the theory with the least number of new representations to cancel the gauge anomalies presented in Ref.~\cite{Perez:2014qfa}. As we will show, it is crucial to consider a complete theory in order to predict the decays of the new Higgs (to which we will refer as Baryonic Higgs) since the loop-induced decays with the anomaly-canceling fermions running in the loop can be the dominant ones due to the fact that they do not suffer from mixing suppression. We show that one can predict the branching ratios for the Baryonic Higgs in agreement with the cosmological constraints on the dark matter density. Our main predictions can be tested in the near future at the LHC and dark matter experiments.

The paper is structured as follows: In Section~\ref{sec:DM} we study the dark matter phenomenology and characterize the properties of the second Higgs in the theory. In Section~\ref{sec:Higgs} we discuss the production of this new Higgs at the LHC and the signatures to look for it. We present our conclusions in Section~\ref{sec:Summary}. In Appendix~\ref{sec:appFR} we present all the Feynman rules that were used in our calculations. Finally, Appendix~\ref{sec:appDecays} contains the full analytical results for all the loop-induced and tree-level decays of the Baryonic Higgs. 

\section{Higgs Decays and DM from Anomaly Cancellation}
\label{sec:DM}
In the previous section we discussed a class of theories for dark matter where the dark matter mass is generated by the spontaneous symmetry breaking of the gauge symmetry. In this context, the dark matter candidate is charged under the new force and, therefore, it is part of the anomalons, i.e. fermions needed to satisfy the anomaly cancellation conditions that otherwise would spoil the gauge invariance of the theory. One of the most attractive examples of this type of theories is local baryon number since
\begin{itemize}
\item[(a)] it is one of the simplest extensions of the SM where it is possible to understand the spontaneous breaking of baryon number without giving issues regarding the stability of the proton, which in these theories is predicted to be stable to any order in perturbation theory,
\item[(b)] it predicts the existence of dark matter from anomaly cancellation,
\item[(c)] the cosmological bound on the relic density measured by the Planck satellite $\Omega h^2 \leq 0.12$ translates into an upper bound of $ M_{Z_B} \lesssim 28$ TeV~\cite{FileviezPerez:2019jju} and since all new particles acquire their mass from the symmetry breaking scale, this implies an upper bound on the full theory,
\item[(d)] this theory can live at the low scale in agreement with all experimental constraints with a gauge coupling of order $g_B \sim 0.1$. Therefore, one can hope to test this theory at the LHC.
\end{itemize}
In this work we will discuss the minimal theory for local baryon number~\cite{Perez:2014qfa}, where only four extra fermionic representations are added to define an anomaly-free theory. These new fermionic fields are given by 
\begin{eqnarray}
&& \Psi_L = \mqty(\Psi_1^+ \\ \Psi_1^0)_L \sim (\mathbf{1}, \mathbf{2}, 1/2, 3/2), \quad \Psi_R = \mqty(\Psi_2^+ \\ \Psi_2^0)_R\sim (\mathbf{1},\mathbf{2},1/2,-3/2), \nonumber \\[2ex]
 && \Sigma_L = \frac{1}{\sqrt{2}}\mqty(\Sigma^0 & \sqrt{2}\Sigma^+ \\ \sqrt{2}\Sigma^- & - \Sigma^0)_L\sim (\mathbf{1}, \mathbf{3}, 0, -3/2), \quad  \text{and} \quad \chi_L^0\sim (\mathbf{1}, \mathbf{1}, 0, -3/2) \nonumber,
 \end{eqnarray}
where the numbers in parenthesis correspond to the quantum numbers under the gauge groups $\SU(3)_c$, $\SU(2)_L$, $\U(1)_Y$ and $\U(1)_B$, respectively.
The Yukawa interactions in this theory are given by
\begin{align}
-\mathcal{L} \supset & \hspace{1mm} y_1\overline{\Psi}_R H \chi_L + y_2 H^{\dagger}\Psi_L \chi_L + y_3 H^{\dagger} \Sigma_L \Psi_L + y_4 \overline{\Psi}_R \Sigma_L H \nonumber\\
&+y_{\Psi} \overline{\Psi}_R \Psi_L S^*_B + \frac{y_{\chi}}{\sqrt{2}}\chi_L \chi_L S_B + y_{\Sigma}\text{Tr}\Sigma^2_L S_B + {\rm h.c. },
\label{eq:LB}
\end{align}
where $H\sim (\mathbf{1},\mathbf{2},1/2,0)$ is the SM Higgs and the new Higgs, whose quantum numbers are indirectly fixed by anomaly cancellation,
\begin{equation}
S_B \sim (\mathbf{1},\mathbf{1},0,3),
\end{equation}
is responsible for the spontaneous breaking of $\U(1)_B$ and generates the masses for the new fermions in the theory, including the dark matter candidate.

The scalar potential of this theory reads as
\begin{align}
	V = -\mu_H^2 H^{\dagger} H + \lambda_H (H^{\dagger} H)^2 - \mu_B^2 S^{*}_B S_B + \lambda_B (S^{*}_B S_B)^2 + \lambda_{HB} (H^{\dagger}H)(S^{*}_B S_B),
	\label{eq:potential}
\end{align}
and the Higgses can be written as
\begin{align}
	H = \mqty(h^+ \\ \frac{1}{\sqrt{2}}(h_0 + i a_0)),\quad  \text{ and } \quad S_B = \frac{1}{\sqrt{2}}(s_B + i a_B).
\end{align}
Spontaneous symmetry breaking of baryon number is achieved once $S_B$ acquires the vev, $\langle S_B\rangle = v_B/\sqrt{2}$, while the electroweak spontaneous symmetry is triggered by the vev of the SM Higgs, $\langle h^0 \rangle = v_0/\sqrt{2}$. Both Higgses mix through the scalar potential in Eq.~\eqref{eq:potential}. In the broken phase, 
the physical Higgses are defined as follows
\begin{equation}
\begin{split}
	h &= h_0 \cos{\theta_B} - s_B \sin{\theta_B},\\
	h_B &= s_B \cos{\theta_B} + h_0 \sin{\theta_B},
	\end{split}
\end{equation}
where the mixing angle $\theta_B$ that diagonalizes the mass matrix for the Higgses is given by
\begin{align}
	\tan{2 \theta_B} = \frac{v_0 v_B \lambda_{HB}}{ v_B^2 \lambda_B - v_0^2 \lambda_H }.
\end{align}
This theory predicts the existence of a new gauge boson $Z_B$, which does not interact with leptons at tree-level, and a Baryonic Higgs $h_B$. For the physical fermion fields we have two charged $F^{\pm}_i  \,\, (i=1,2)$ and four neutral: $F^0_j \,\, (j=1,2,3)$ and the dark matter candidate $\chi$.

After spontaneous symmetry breaking, the local $\U(1)_B$ is broken to a ${\mathcal{Z}}_2$ symmetry which protects the dark matter candidate from decaying. This symmetry acts only in the new sector as follows: 
$$\{ \Psi_L \to - \Psi_L , \ \Psi_R \to - \Psi_R, \  \Sigma_L \to - \Sigma_L , \ \chi_L^0 \to - \chi_L^0 \}.$$
In order to have a consistent scenario for cosmology, we then require that $$M_{\chi} < M_{F_i^0}, M_{F^{\pm}_i },$$ and the lightest neutral state can be a good dark matter candidate because it is automatically stable and neutral.
In this article we will investigate the dark matter properties when $\chi \sim \chi_L +( \chi_L)^C$ since in this case the Baryonic Higgs can have a large branching ratio to dark matter even when its mass is not far from the electroweak scale. For simplicity, we will take the limit where the Yukawa couplings with the SM Higgs $y_i$, with $i=1, ..., 4$, are negligible. In this context, the Feynman rules of the theory are listed in Appendix~\ref{sec:appFR}.

The most relevant parameters for our dark matter studies are the new gauge coupling, $g_B$, the mixing angle between the two Higgses, $\theta_B$, the dark matter mass, $M_\chi$, the mass of the Baryonic Higgs, $M_{h_B}$, and the mass of the new gauge boson, $M_{Z_B}$. The masses of the anomaly-canceling fermions, $M_{F_i^0} \text{ and } M_{F^{\pm}_i }$, become relevant when we study the Baryonic Higgs decays.  
For phenomenological studies of this type of theory see Refs.~\cite{Ohmer:2015lxa,Duerr:2017whl,Duerr:2017uap,Duerr:2014wra,FileviezPerez:2020mtk,FileviezPerez:2019jju,FileviezPerez:2018jmr,FileviezPerez:2020gfb}.

\subsection{Dark Matter: Relic Density and Direct Detection}
%
In order to investigate the predictions for the Higgs decays we first need to discuss the dark matter relic density and direct detection constraints. Since we are mainly interested in the scenarios with a light Baryonic Higgs $h_B$, so that it can be produced at the LHC, we focus on the simplest scenario where $\chi$ is singlet-like under the SM. 
In the context of the minimal theory for baryon number described in the previous section, our dark matter candidate is a Majorana fermion, $\chi=\chi_L + (\chi_L)^C$, and has
the following annihilation channels,
$$\chi \chi \to \bar{q}q, \, Z_B Z_B, \, Z_B h, \, Z_B h_B, \, h h, \, h h_B, \, h_B h_B, \, WW, \, ZZ.$$
Notice that the channels $Z_B h, \, h h, \, h h_B, \, WW$ and $ZZ$ are suppressed by the mixing angle $\theta_B$, which from collider bounds has to be $\sin \theta_B \leq 0.3$~\cite{Ilnicka:2018def,Adhikari:2020vqo}.
The channels $\bar{q}q, \, Z_B Z_B, Z_B h_B$ and $h_B h_B$, if allowed, do not suffer from mixing suppression and also define the most interesting regions that satisfy the relic density constraints, as we show in this section.
\begin{figure}[h]
\centering
\includegraphics[width=0.495\linewidth]{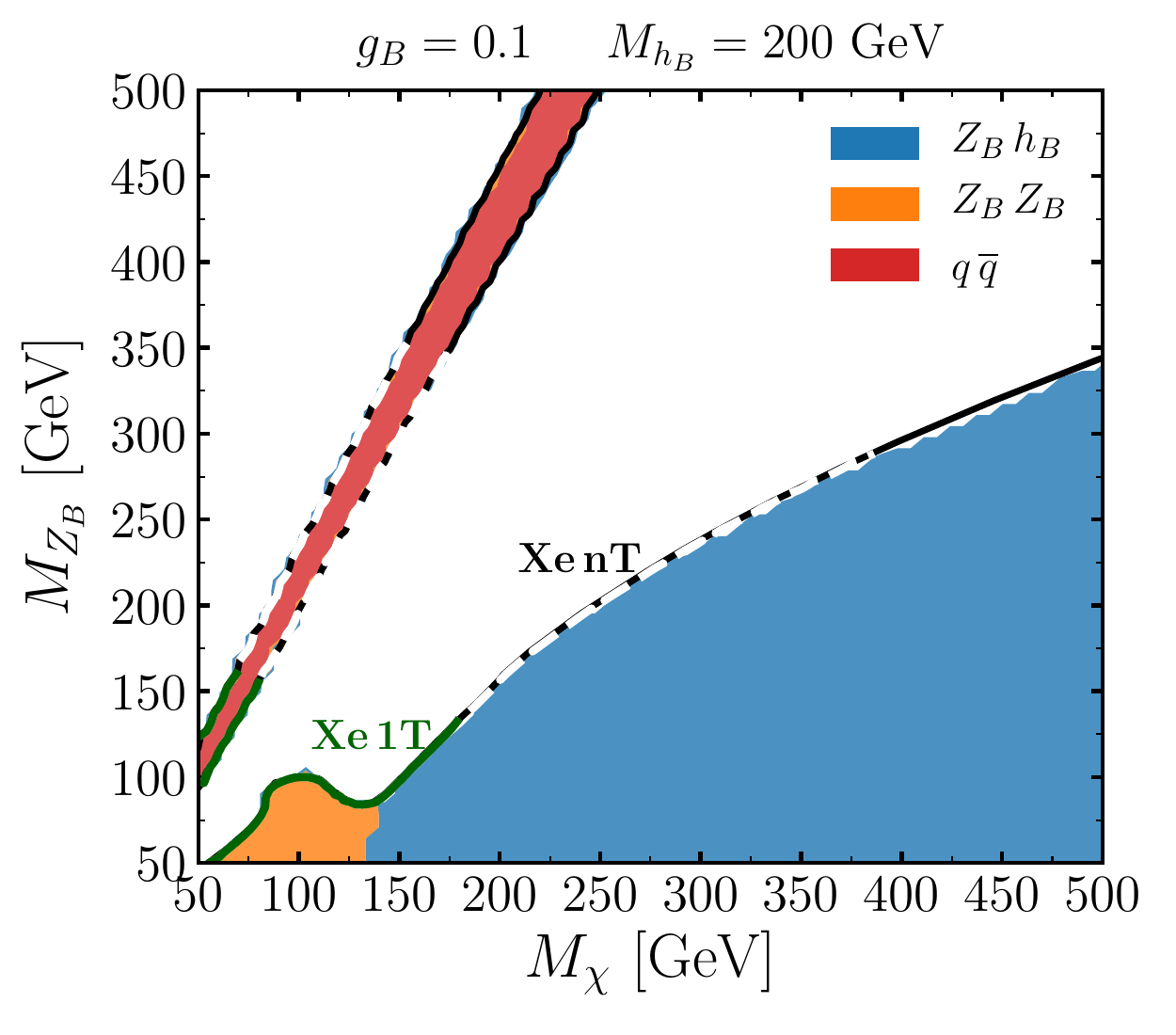} 
\includegraphics[width=0.495\linewidth]{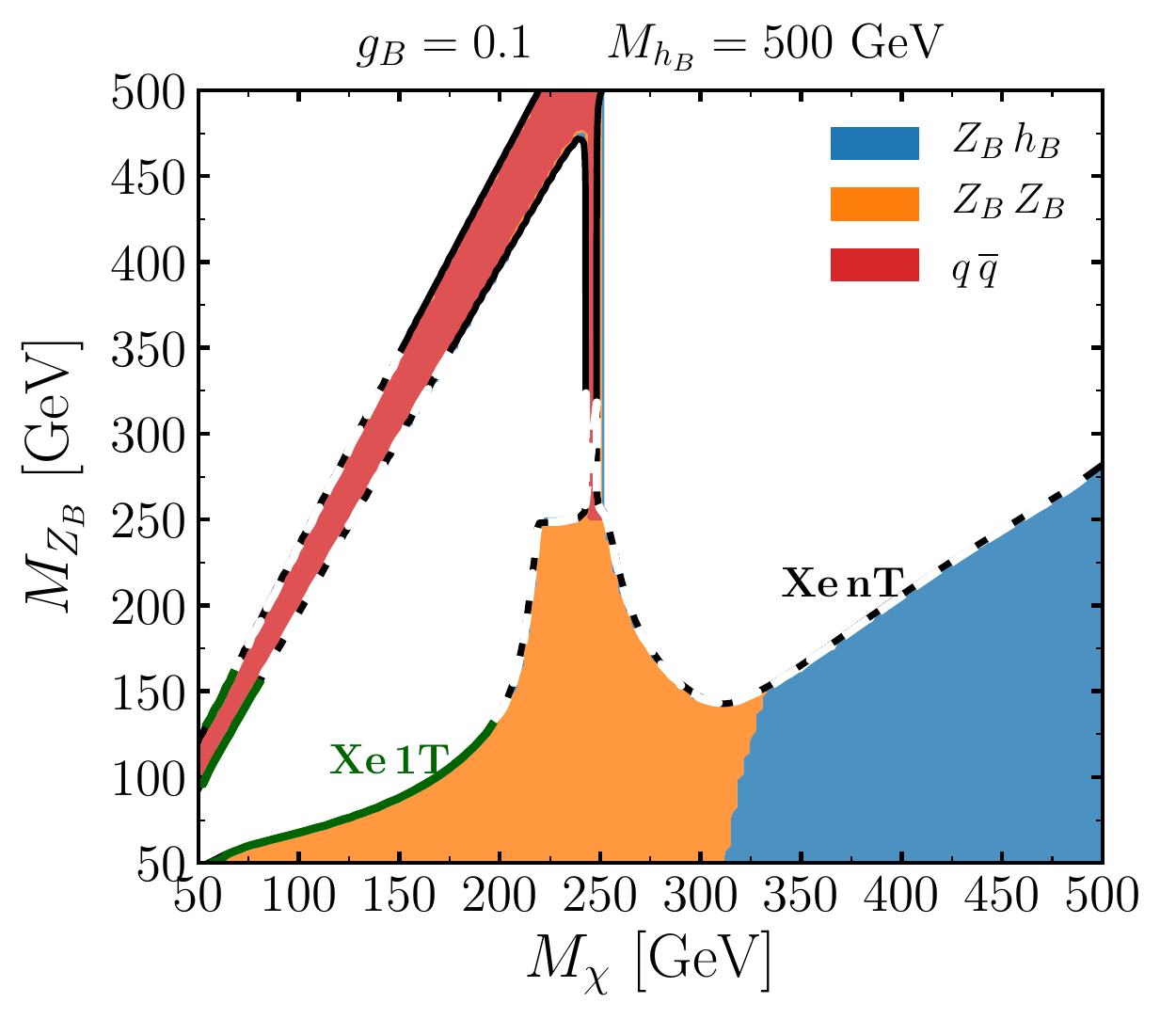} 
\caption{Contour plot for the dark matter relic density in the $M_\chi$ versus $M_{Z_{B}}$ plane. In the left (right) panel the mass of the second Higgs has been fixed to $M_{h_B}=200$ GeV (500 GeV).  The solid black line gives the measured relic abundance by the Planck satellite $\Omega_\chi h^2 =0.1200 \pm 0.0012$~\cite{Aghanim:2018eyx}. The colored region satisfies the constraint $\Omega_\chi h^2 \leq 0.12$ and the different colors correspond to different dominant annihilation channels as indiciated in the legend. The solid green line shows the region ruled out by Xenon-1T \cite{Aprile:2018dbl}, the dashed black line shows the projected sensitivity for Xenon-nT \cite{Aprile:2015uzo}, both for zero mixing angle. For small mixing angles, the dark matter relic density and the direct detection are almost independent of the scalar mixing angle as explained in the text.}
\label{fig:relic1}
\end{figure}

In Fig.~\ref{fig:relic1} we present our results for the calculation of the dark matter relic density which has been computed numerically using \texttt{MicrOMEGAs 5.0.6} \cite{Belanger:2018ccd}. 
In both panels we show, in solid black, the contour lines that give the relic abundance measured by the Planck satellite $\Omega_\chi h^2 =0.1200 \pm 0.0012$~\cite{Aghanim:2018eyx}. The colored regions show the parameter space that gives a relic density that agrees with the later bound but lies below it. On the left panel, the mass of the Baryonic Higgs is fixed to $M_{h_B} = 200$ GeV, while on the right panel we show the parameter space in agreement with the relic density bound for a heavier Baryonic Higgs of $M_{h_B} = 500$ GeV. For these plots, as well as for the rest of the results presented here, we will assume a gauge coupling $g_B = 0.1$ in order to be in agreement with the collider bounds coming from dijet searches regardless of the $Z_B$ mass considered. The monojet bounds become relevant ony for smaller masses below the range we consider~\cite{ATLAS:2020wzf}. For a detailed study of the dark matter phenomenology for larger values of the gauge coupling see Ref.~\cite{FileviezPerez:2020mtk}.

\begin{table}
\centering
\begin{tabular}{ | c | l | c|}
\hline
\rowcolor{GrayLight}
Label & Dominant annihilation channel & Properties \\
\hline
Scenario I & \quad \quad \quad \quad   $\chi \chi \to \bar q q$ & $\! M_\chi \simeq M_{Z_B}/2$ \\
Scenario II &  \quad \quad \quad \quad $ \chi \chi \to Z_B h_B$ & $\quad \quad \quad 2 M_\chi > M_{h_B}+M_{Z_B}$ \\
Scenario III &  \quad \quad \quad \quad $ \chi \chi \to Z_B Z_B$ & $\!\!\! \! M_\chi > M_{Z_B}, \,\,\,\,\, M_\chi \simeq M_{h_B}/2$  \\
\hline
\end{tabular}
\caption{Classification of the three different scenarios  for which the relic density bound $\Omega_\chi h^2 =0.1200 \pm 0.0012$~\cite{Aghanim:2018eyx} is satisfied in the minimal theory for local baryon number. }
\label{Scenarios}
\end{table}

The parameter space of the theory can be classified depending on which annihilation channel gives the largest contribution to the relic density, as illustrated in Table~\ref{Scenarios}, what gives rise to four possible scenarios:
\begin{itemize}
\item{ {\bf Scenario I} ($\chi \chi \to \bar{q}q$)}: The annihilation into a quark antiquark pair is the dominant one only at the resonance $M_\chi \approx M_{Z_B}/2$. This region is shown by the thin diagonal in Fig.~\ref{fig:relic1}. This region satisfies the relic density bound independently of the value of $M_{h_B}$. 
\item{{\bf Scenario II} ($\chi \chi \to Z_B h_B$)}: As shown in both plots in Fig.~\ref{fig:relic1} the $\chi \chi \to Z_B h_B$ annihilation can be dominant in a large region of the parameter space and it does not rely on a resonance.
\item{{\bf Scenario III} ($\chi \chi \to Z_B Z_B$)}: In the scenario with $M_{h_B} \gg M_\chi$ we have to rely on the $Z_B Z_B$ annihilation channel to achieve the measured relic abundance. As shown on the left plot in Fig.~\ref{fig:relic1}, this annihilation can be dominant in the resonance $M_\chi \approx M_{h_B}/2$. However, it does not always have to be close to the resonance, as shown in the right plot of Fig.~\ref{fig:relic1}. This scenario, while disfavoured by direct detection searches for light $h_B$ masses, will be fully probed by Xenon-nT in the near future for $M_{h_B} \lesssim 1 \text{ TeV}$. 
\item{{\bf Scenario IV} ($\chi \chi \to h_B h_B)$}: This annihilation is dominant in the limit when $M_{h_B} \ll M_\chi \ll M_{Z_B}$ and a large Yukawa coupling $y_\chi \approx M_\chi/v_B$. These conditions imply a large gauge coupling above the perturbativity limit so we do not discuss this scenario any further.
\end{itemize}
%
%
In Fig.~\ref{fig:relic1}, we color each region depending on which annihilation channel gives the dominant contribution to the relic abundance: the red area is dominated by $\chi \chi \to \bar q q$, the orange area is dominated by $\chi \chi \to Z_B Z_B$, while in the blue area the annihilation channel $\chi \chi \to Z_B h_B$ dominates.

Regarding direct detection of the dark matter, the cross-section mediated by the $Z_B$ is velocity suppressed,
\begin{eqnarray}
\sigma_{\chi N}^\text{SI}(Z_B)&=& \frac{27}{8 \pi}\frac{g_B^4 M_N^2}{M_{Z_B}^4} v^2,
\end{eqnarray}
where $M_N$ is the nucleon mass and $v$ the velocity of the non-relativistic dark matter. There is also the channel mediated by Higgs mixing:
\begin{equation}
\sigma_{\chi N}^\text{SI}(h_i)=\frac{72 }{8 \pi v_0^2}\sin^2 \theta_B \cos^2 \theta_B M_N^4 \frac{g_B^2 M_\chi^2}{M_{Z_B}^2}\left(\frac{1}{M_{h}^2}-\frac{1}{M_{h_B}^2}\right)^2 f_N^2,
\end{equation}
which is suppressed by the mixing angle. The parameter $f_N$ corresponds to the effective Higgs-nucleon-nucleon coupling that we take $f_N=0.3$ \cite{Hoferichter:2017olk}.

\begin{figure}[h]
\centering
\includegraphics[width=0.65\linewidth]{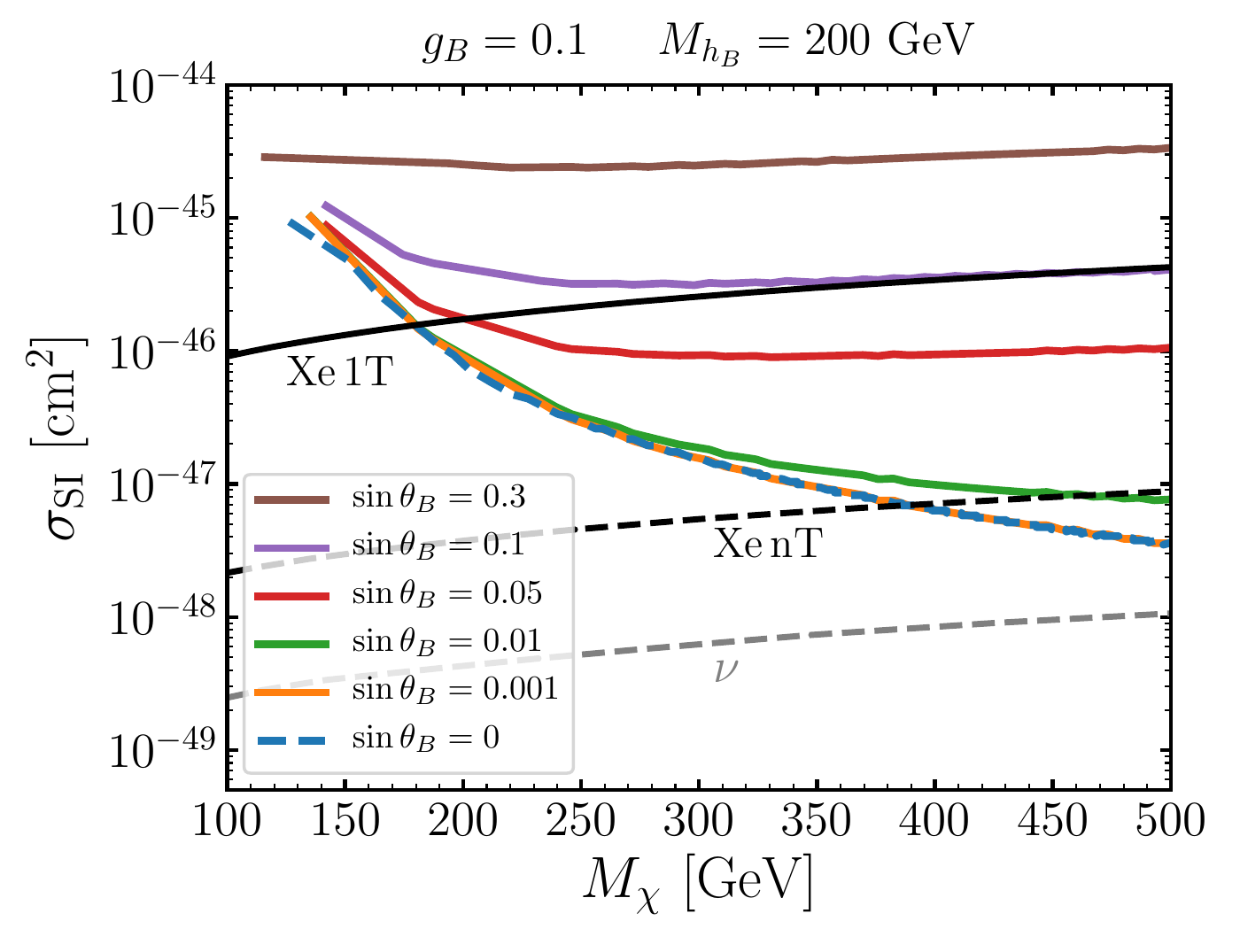} 
\caption{Predictions for the direct-detection spin-independent cross-section as a function of the dark matter mass. The purple, red, green and orange lines correspond to $\sin \theta_B = 0.1$, $0.05$, $0.01$ and $0.001$, respectively. The brown line corresponds to the maximal mixing scenario, $\sin \theta_B = 0.3$, while the blue dashed line represents the minimal one, $\theta_B=0$. All points give the measured relic abundance by the Planck satellite $\Omega_\chi h^2 =0.1200 \pm 0.0012$~\cite{Aghanim:2018eyx} and satisfy constraints from the LHC. The solid black lines show current experimental bounds from Xenon-1T~\cite{Aprile:2017iyp, Aprile:2018dbl}, the dashed black line shows the projected sensitivity for Xenon-nT~\cite{Aprile:2015uzo} and the dashed gray line shows the coherent neutrino scattering limit \cite{Billard:2013qya}.}
\label{DD}
\end{figure}

In Fig.~\ref{fig:relic1} we show in solid green the parameter space ruled out by Xenon-1T~\cite{Aprile:2017iyp, Aprile:2018dbl}, while the dashed black line shows the region that will be probed by Xenon-nT~\cite{Aprile:2015uzo} in the case of zero mixing angle, these constraints become stronger for larger mixing angles.  In Fig.~\ref{DD} we present our predictions for the spin-independent cross-section as a function of the dark matter mass with all points satisfying the measured relic abundance. As can be seen, the projected sensitivity for Xenon-nT will probe the zero mixing angle case for $M_{\chi}<390$ GeV. Strikingly, this result tells us that one could test this theory if our dark matter candidate is not too heavy. 

\subsection{Baryonic Higgs Decays}
The new Higgs present in the theory can have the following decays:
$$h_B \to \gamma \gamma,\,\, gg, \,\, \gamma Z, \,\, \gamma Z_B, \,\, ZZ, \,\, Z Z_B, \,\, Z_B Z_B, \,\, WW, \,\, { \chi} \chi, \,\, \overline{F_i} F_i, \,\, hh, \,\, \overline{f_i} f_i, $$
where $\chi$, $F_i$ and $f_i$ correspond to dark matter, anomaly-canceling fermions (also referred to as anomalons throughout the text) and SM fermions, respectively. All the Feynman diagrams and the full analytical results for the tree-level and loop-induced decays of the Baryonic Higgs are given in Appendix~\ref{sec:appDecays}.
Among them, the decays $h_B \to Z_B Z_B,  \,\, \chi \chi,  \,\, \bar{F} F$ occur at tree-level, with no suppression if allowed by kinematics. Therefore, when any of these decays is open, it dominates the branching ratios. The decays $h_B \to \gamma \gamma, \,\, Z \gamma$ are loop suppressed but can be dominant whenever the tree-level decays are kinematically closed. The decays $h_B \to \,\, WW, \,\, ZZ$  have a tree-level and one-loop contribution, the former is the dominant for large mixing angles and the latter dominates for small mixing angles. The rest of decays, $h_B \to Z_B \gamma, \,\, gg, \,\, ZZ_B$, on top of being loop-induced are suppressed by the mixing angle $\theta_B$ as well. In the limit of very heavy anomaly-canceling fermions these decays were studied in Ref.~\cite{Ohmer:2015lxa}.

As we discussed in the previous sections, having a complete theory that includes the anomaly-canceling fermions is crucial to make the correct predictions for the different decays of the Baryonic Higgs. In the previous section we demonstrated that the scalar mixing angle $\theta_B$  is strongly constrained by direct detection bounds, which has the following implications for the Baryonic Higgs decays in the three different scenarios that can be in agreement with the cosmological bound on the relic abundance:
\begin{itemize}
\item{{\bf Scenario I }($\chi \chi \to \bar{q}q$)}: In this scenario, the decays $h_B \to \chi \chi$ and $h_B \to Z_B Z_B$ can be open or closed depending on the value of $M_{h_B}$. The left panel in Fig.~\ref{fig:BRqq} shows the results for $ M_{h_B} < 2 M_\chi < 2 M_{Z_B}$, where the invisible decays and the decay to a pair of new gauge bosons are kinematically closed, while in the right panel we show the results for $2 M_\chi < M_{h_B} < 2 M_{Z_B}$, and hence, $h_B$ decays predominantly into dark matter.
\begin{figure}[h!]
\centering
\includegraphics[width=0.49\linewidth]{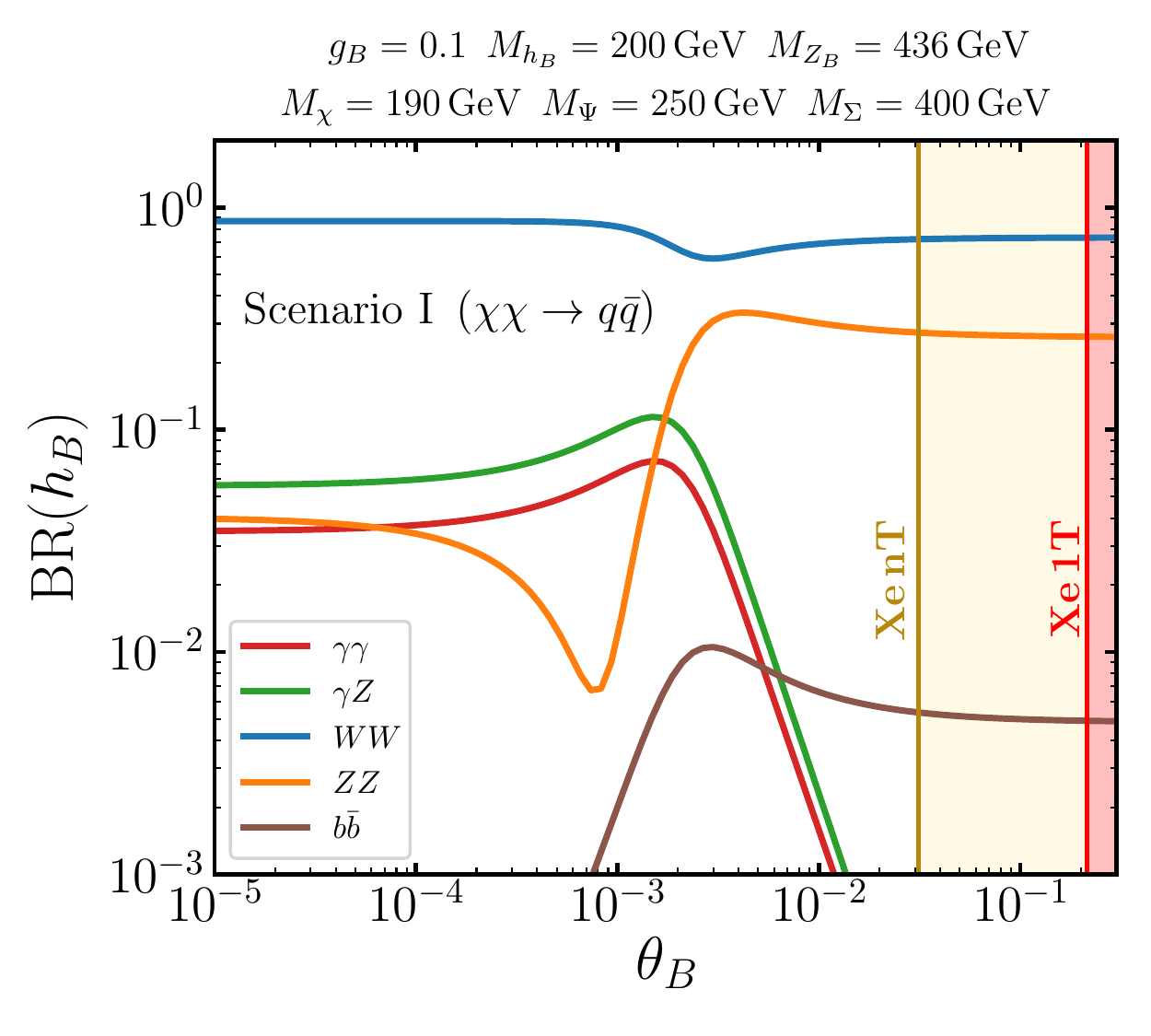}
\includegraphics[width=0.49\linewidth]{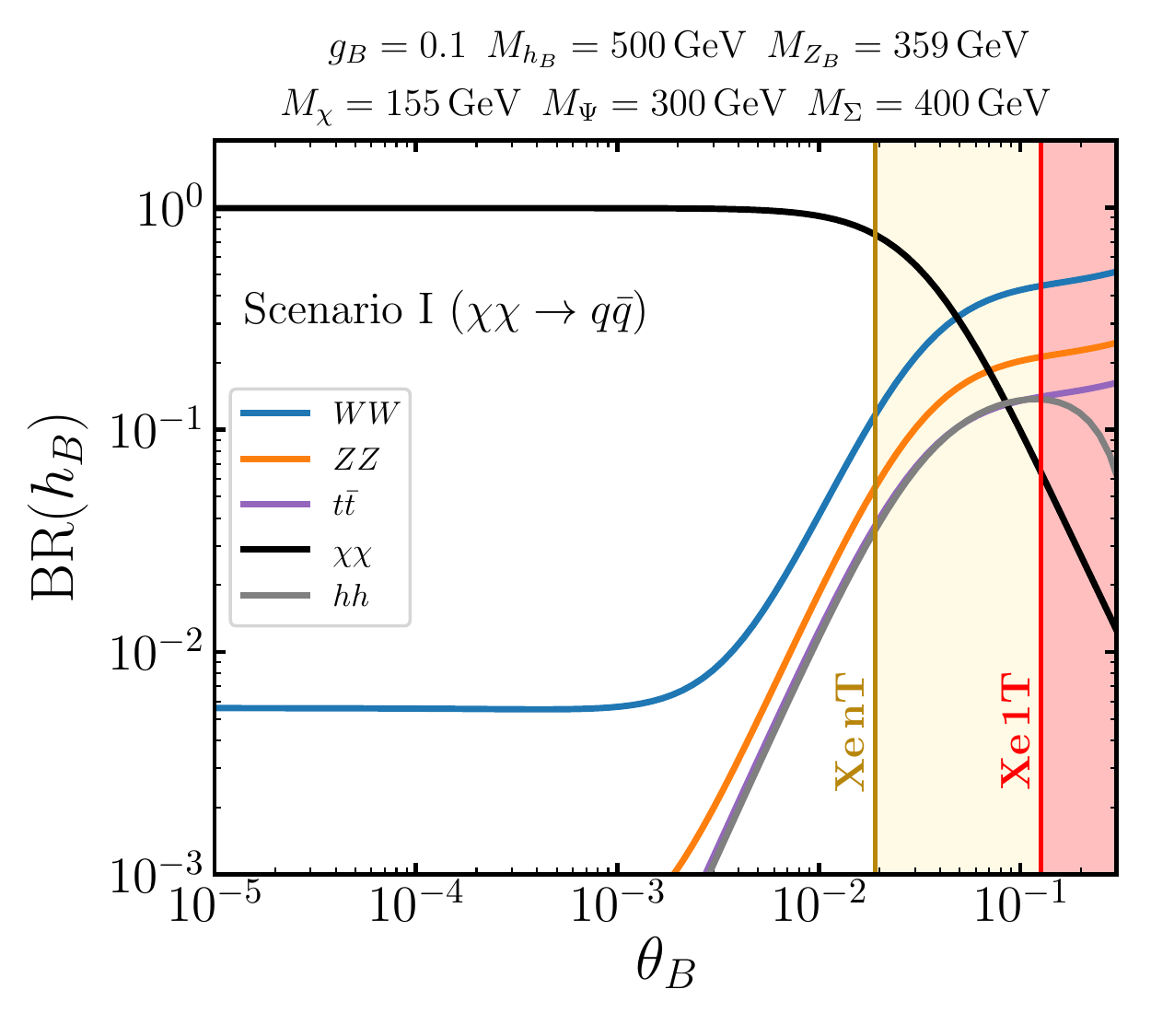} 
\caption{Branching ratios of the Baryonic Higgs $h_B$ as a function of the scalar mixing angle $\theta_B$ in the context of Scenario I, where $\chi \chi \to \bar q q$ is the dominant annihilation channel. The relevant parameters have been fixed as indicated in the plot to satisfy the dark matter relic density $\Omega h^2 = 0.12$. The different colors correspond to different decay channels as shown in the plot. The area shaded in red is ruled out by Xenon-1T direct detection constraints, while in yellow we show the region that will be probed by Xenon-nT.}
\label{fig:BRqq}
\end{figure}
\item{{\bf Scenario II }($\chi \chi \to Z_B h_B$)}: This scenario requires that $2M_\chi > M_{h_B} +M_{Z_B}$, and therefore, the invisible decay is closed. Furthermore, $M_
{Z_B}$ has to be around $M_\chi$ for low masses of the dark matter in order to achieve the correct relic density, so that in this case the $h_B \to Z_B Z_B$ decay is also closed. This implies that for small mixing angles the loop-induced decays into $\gamma \gamma$ and $\gamma Z$ can be significant, as it is shown on the left panel of Fig.~\ref{fig:BRZBhB}. However, as we show in the right panel of that figure, for larger dark matter masses $M_{Z_B} \lesssim M_{\chi}$ and hence the $h_B \to Z_B Z_B$ can dominate for small mixing angles.
\begin{figure}[h!]
\centering
\includegraphics[width=0.49\linewidth]{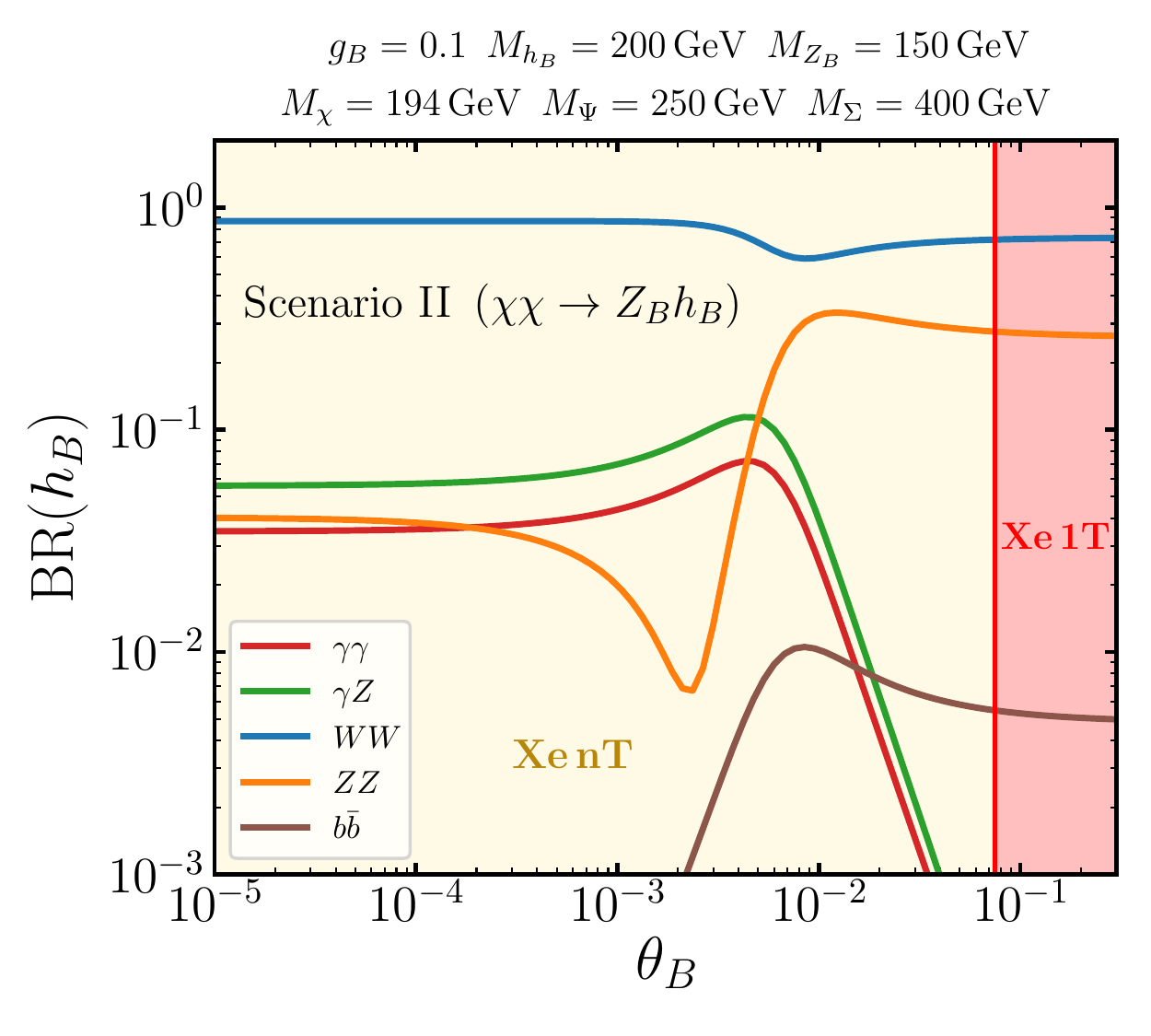} 
\includegraphics[width=0.49\linewidth]{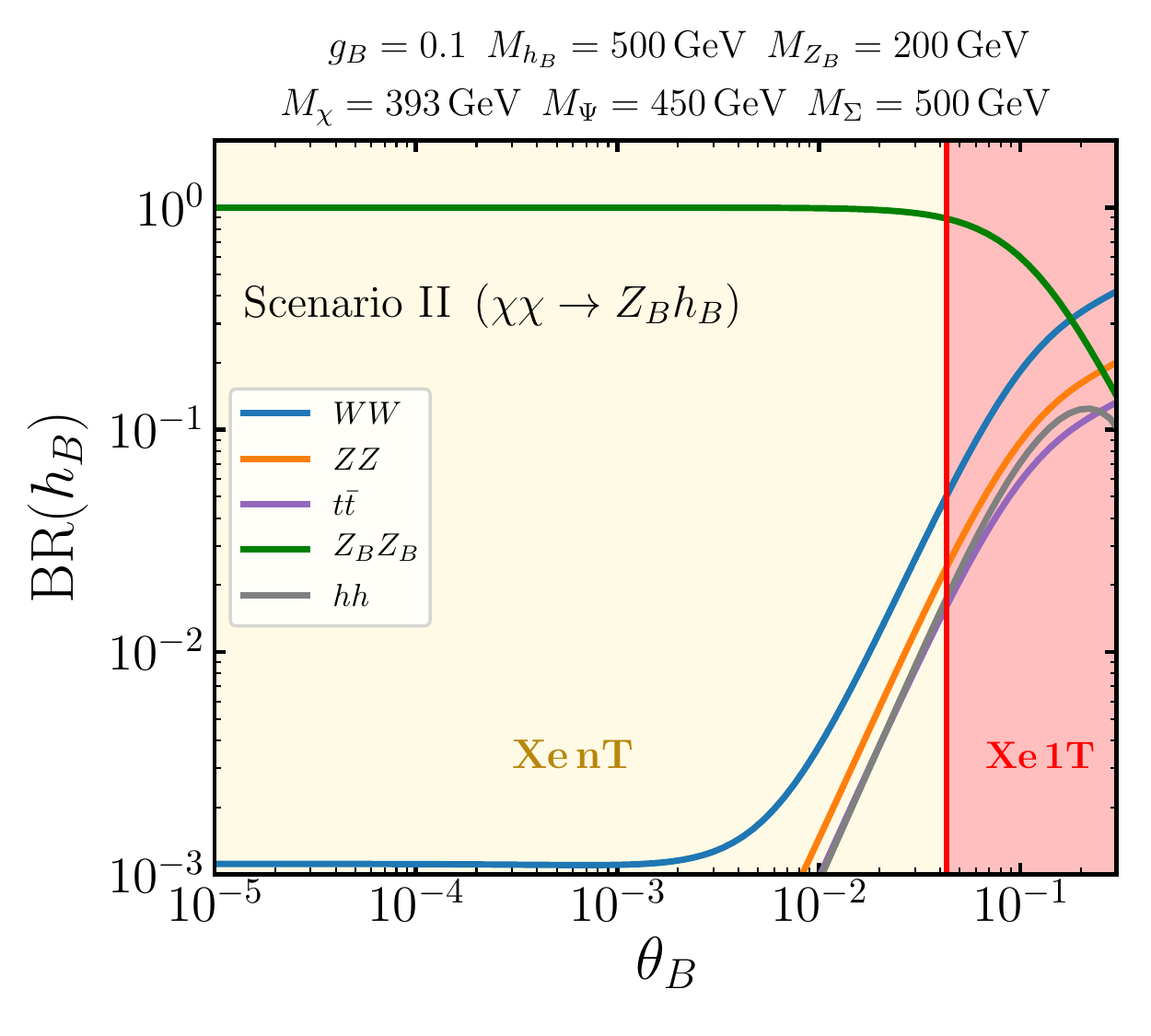} 
\caption{Branching ratios of the Baryonic Higgs $h_B$ as a function of the scalar mixing angle $\theta_B$ in the context of Scenario II, where $\chi \chi \to Z_B h_B$ is the dominant annihilation channel. The relevant parameters have been fixed as indicated in the plot to satisfy the dark matter relic density $\Omega h^2 = 0.12$. The different colors correspond to different decay channels as shown in the plot. The area shaded in red is ruled out by Xenon-1T direct detection constraints, while in yellow we show the region that will be probed by Xenon-nT.}
\label{fig:BRZBhB}
\end{figure}
\item{{\bf Scenario III }($\chi \chi \to Z_B Z_B$)}:
For this annihilation channel to be open, $M_{Z_B} \lesssim M_\chi$ is required. Furthermore, as shown in Fig.~\ref{fig:relic1}, the condition $M_{h_B} \gtrsim 2 M_{Z_B}$ is always satisfied. This implies that the decay $h_B \to Z_B Z_B$ is open and it is the dominant decay. On the other hand, whenever $M_{h_B} \gtrsim 2 M_\chi$ the Higgs will also have a branching ratio into invisible. The branching ratios for the second Higgs in this scenario are shown in Fig.~\ref{fig:BRZBZB}. 
%
\begin{figure}[h!]
\centering
\includegraphics[width=0.49\linewidth]{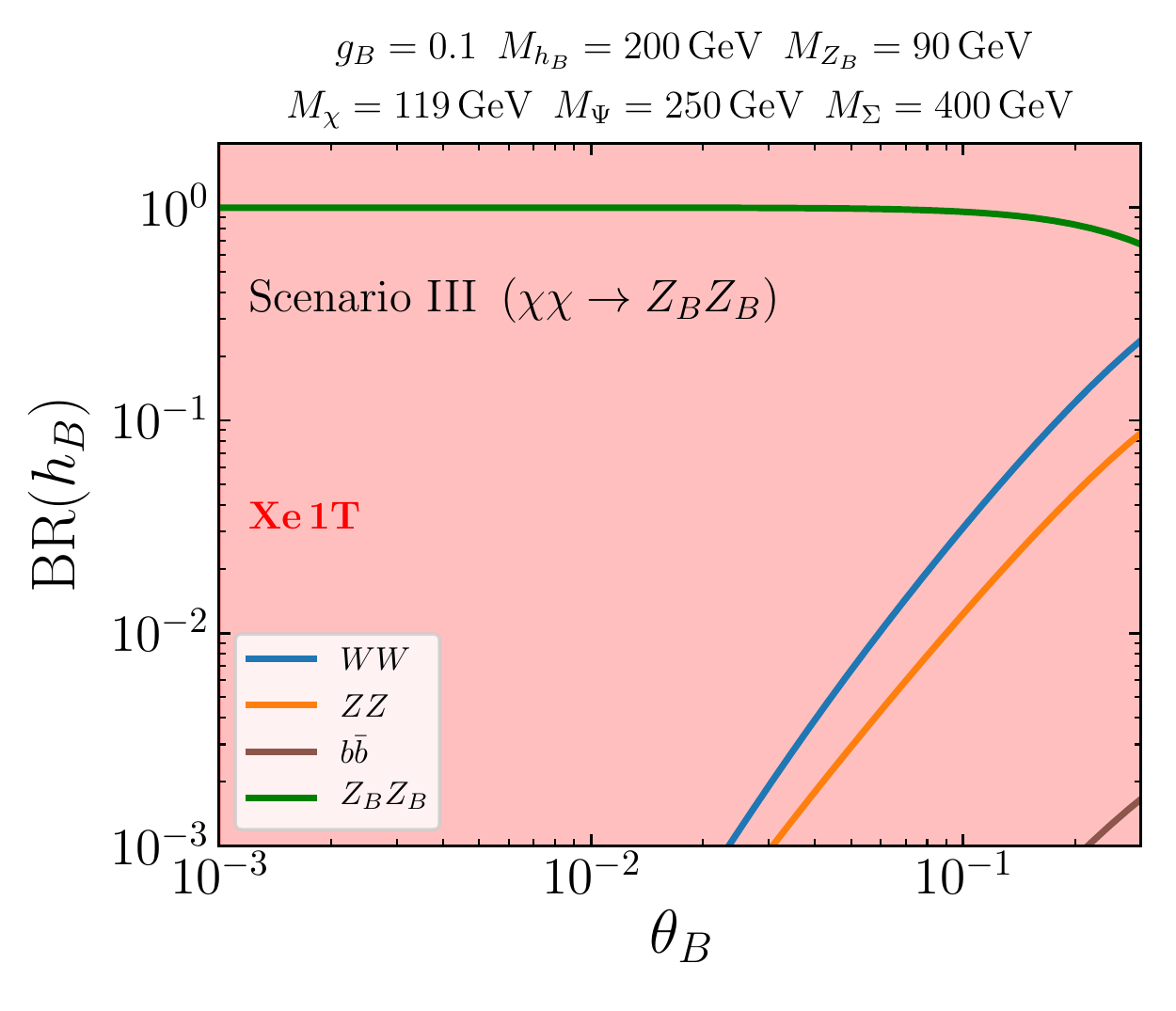} 
\includegraphics[width=0.49\linewidth]{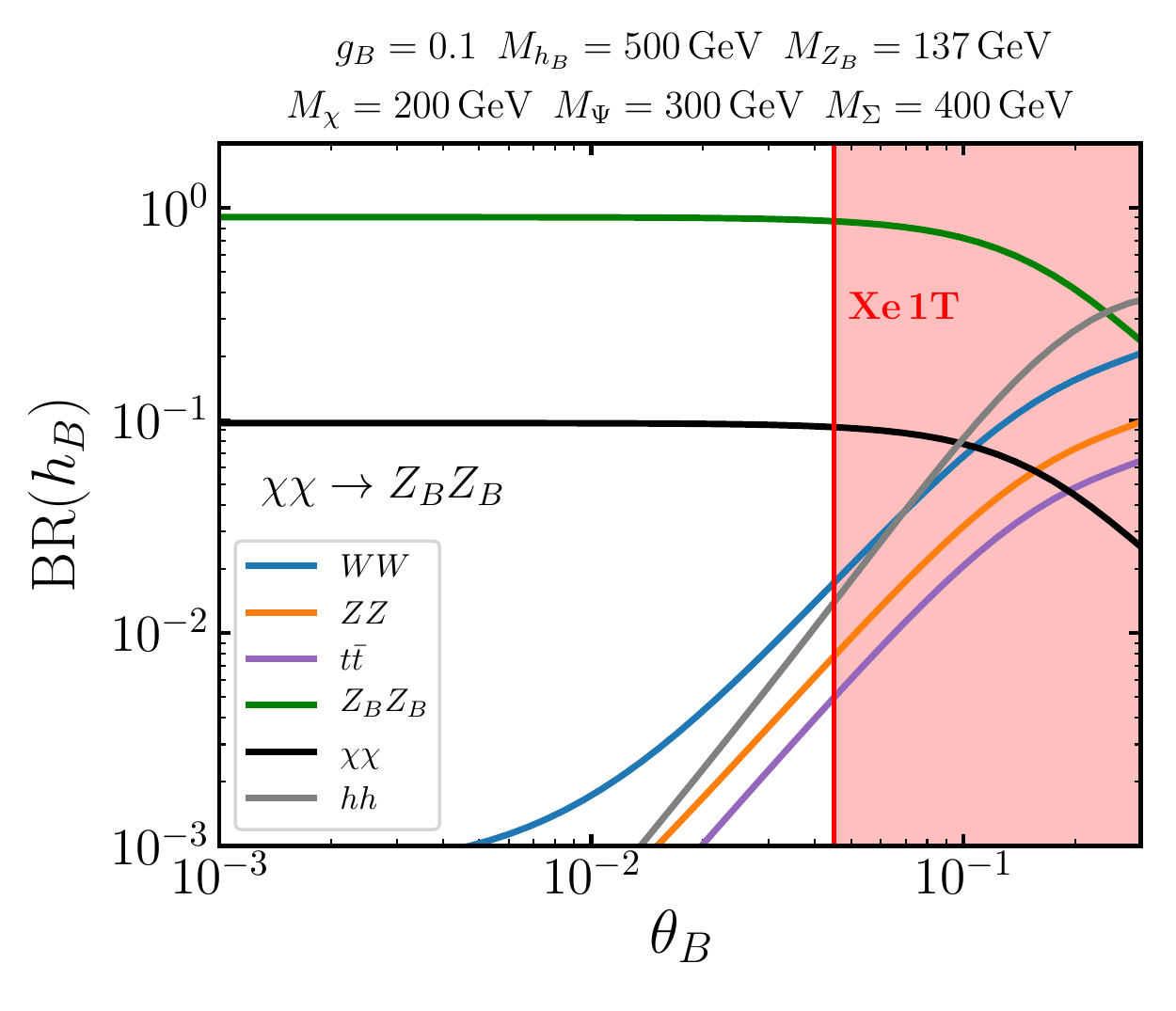} 
\caption{Branching ratios of the Baryonic Higgs $h_B$ as a function of the scalar mixing angle $\theta_B$ in the context of Scenario III, where $\chi \chi \to Z_B Z_B$ is the dominant annihilation channel. The relevant parameters have been fixed as indicated in the plot to satisfy the dark matter relic density $\Omega h^2 = 0.12$. The different colors correspond to different decay channels as shown in the plot. The area shaded in red is ruled out by Xenon-1T.}
\label{fig:BRZBZB}
\end{figure}
\end{itemize}

We would like to mention in passing that for $M_{Z_B}<125$ GeV the SM Higgs can also decay into $h \to \gamma Z_B$ with the top quark in the loop. However, we find that $ {\rm BR}(h \to \gamma Z_B) < 10^{-5}$ which means that this exotic Higgs decay will be hard to detect in the near future.

From the above scenarios, we would also like to highlight the interesting framework of Scenario II where it is possible to have large branching ratios of the Baryonic Higgs to the electroweak gauge bosons and the photon, $\text{BR}(h_B \to \gamma \gamma) \gtrsim 1 \%$. In this case, one can expect to detect this Baryonic Higgs at colliders through the number of events predicted for those clean channels.


\section{Signatures at the LHC}
\label{sec:Higgs}
In this section we study the main signatures of the minimal theory for local baryon number at the LHC. We first focus on the associated production of the Baryonic Higgs and use the fact that its decay to photons or missing energy can be relevant for certain regions of the parameter space. We also highlight that this theory predicts unequivocally the branching ratios of the leptophobic gauge boson $Z_B$.

\subsection{Associated production of $h_B$ and $Z_B$}
We have shown in the previous section that the dark matter direct detection experimental bounds set a strong limit on the mixing between the SM Higgs and the Baryonic Higgs, $\sin \theta_B < 0.1$, which becomes stronger for light dark matter masses, being able to rule out any scalar mixing for $M_\chi \lesssim 180$ GeV for $g_B = 0.1$ as Fig.~\ref{DD} shows. Therefore, all the SM-like Higgs production mechanisms for the Baryonic Higgs are typically suppressed. 

\begin{figure}[h]
\centering
\includegraphics[width=0.5\linewidth]{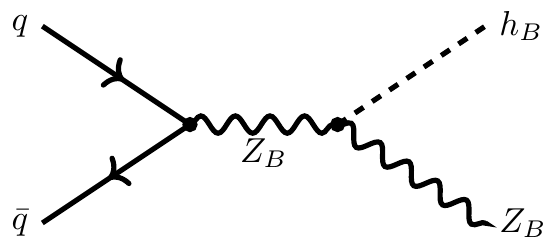}
\caption{Feynman graph for the process $\bar q q \to Z_B^* \to Z_B h_B$.}
\label{fig:FD}
\end{figure}

In these theories we have the possibility to use the associated production mechanism 
$p p \to Z_B^* \to h_B Z_B$, that is not suppressed by the mixing between the two Higgses. 
The cross-section for $pp \to Z_B^* \to Z_B h_B$ is given by
\begin{equation}
\sigma (pp \to h_B Z_B) (s) = \int_{\tau_0}^{1} d\tau \frac{d {\cal L}^{pp}_{q\bar q}}{d\tau}\sigma(q \bar q \to Z_B h_B) (\hat{s}),
\label{eq:hadronicXsection}
\end{equation}
which weights the contribution of the partonic cross-section of the process represented in Fig.~\ref{fig:FD},
\begin{equation}
\sigma = \frac{\cos^2 \theta_B g_B^4 }{144 \pi s^2}\frac{(M_{h_B}^4\! \! - \! 2M_{h_B}^2(M_{Z_B}^2 \!\! + \! s)+M_{Z_B}^4 \! + \! 10M_{Z_B}^2s \! + \! s^2) \sqrt{(s \! + \! M_{Z_B}^2 \!\! - \! M_{h_B}^2)^2 \! - \! 4sM_{Z_B}^2}}{(s \! - \! M_{Z_B}^2)^2 \! + \! \Gamma_{Z_B}^2 M_{Z_B}^2},
\end{equation}
by the contribution of each quark to the proton, the latter being parametrized by the corresponding parton distribution function as follows,
\begin{equation}
\frac{d {\cal L}_{q\bar q}^{pp}}{d\tau} = \int_\tau^1 \frac{dx}{x} \left [f_{q/p}(x,\mu) \, f_{\bar{q}/p} \left(\frac{\tau}{x},\mu \right) + f_{q/p}\left(\frac{\tau}{x},\mu \right) f_{\bar q/p}(x,\mu) \right], 
\end{equation}
where $\tau = \hat{s}/s$, being $\hat{s}$ the partonic center-of-mass energy squared, $s$ is the center-of-mass energy squared at the hadronic level, $\tau_0 = (M_{h_B}+M_{Z_B})^2/s$ is the production threshold, and $\mu$ refers to the factorization scale, which we take it to be $\mu = (M_{h_B} + M_{Z_B})/2$ in our calculations. 
%
\begin{figure}[h]
\centering
\includegraphics[width=0.65\linewidth]{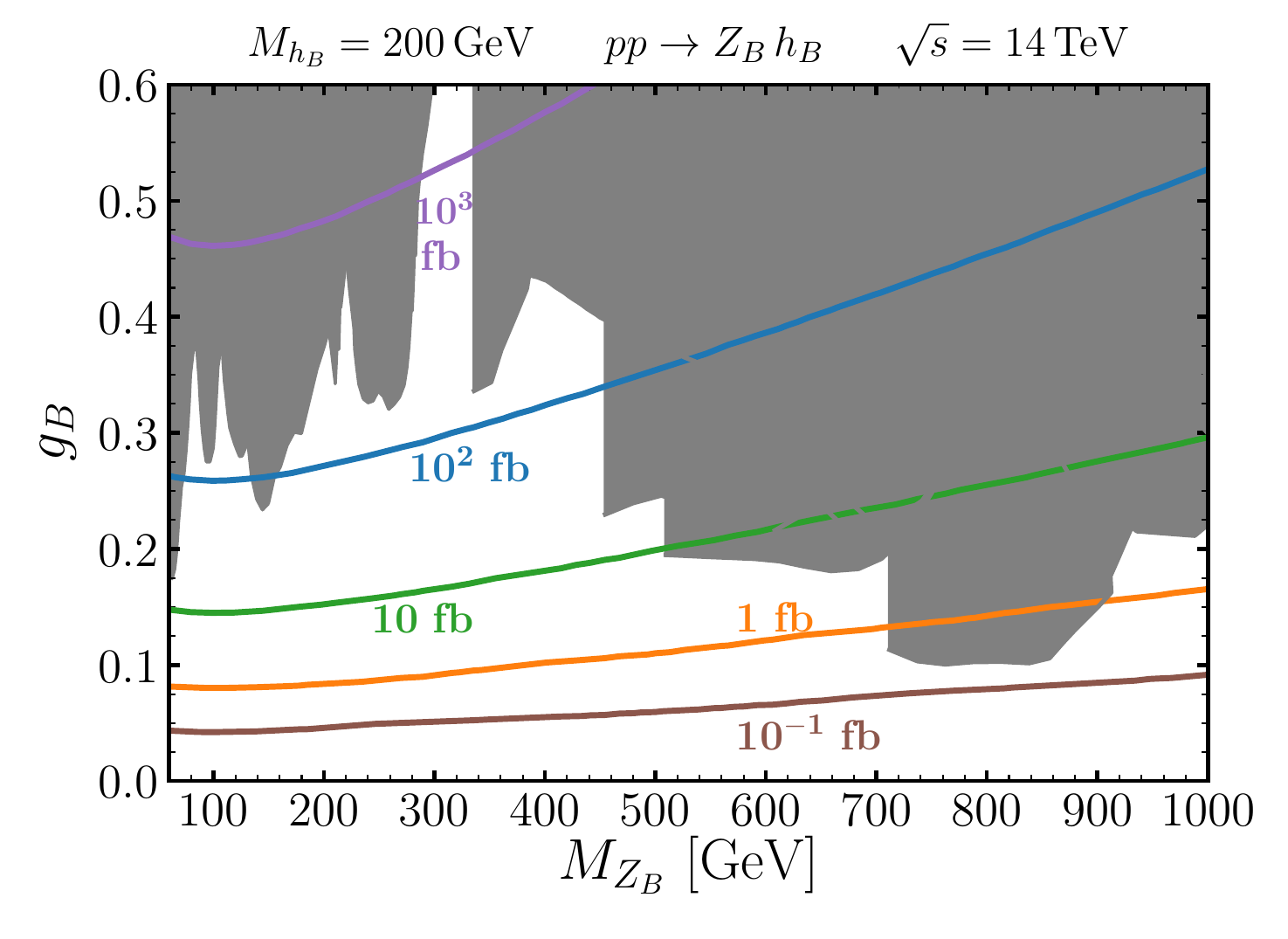}
\includegraphics[width=0.65\linewidth]{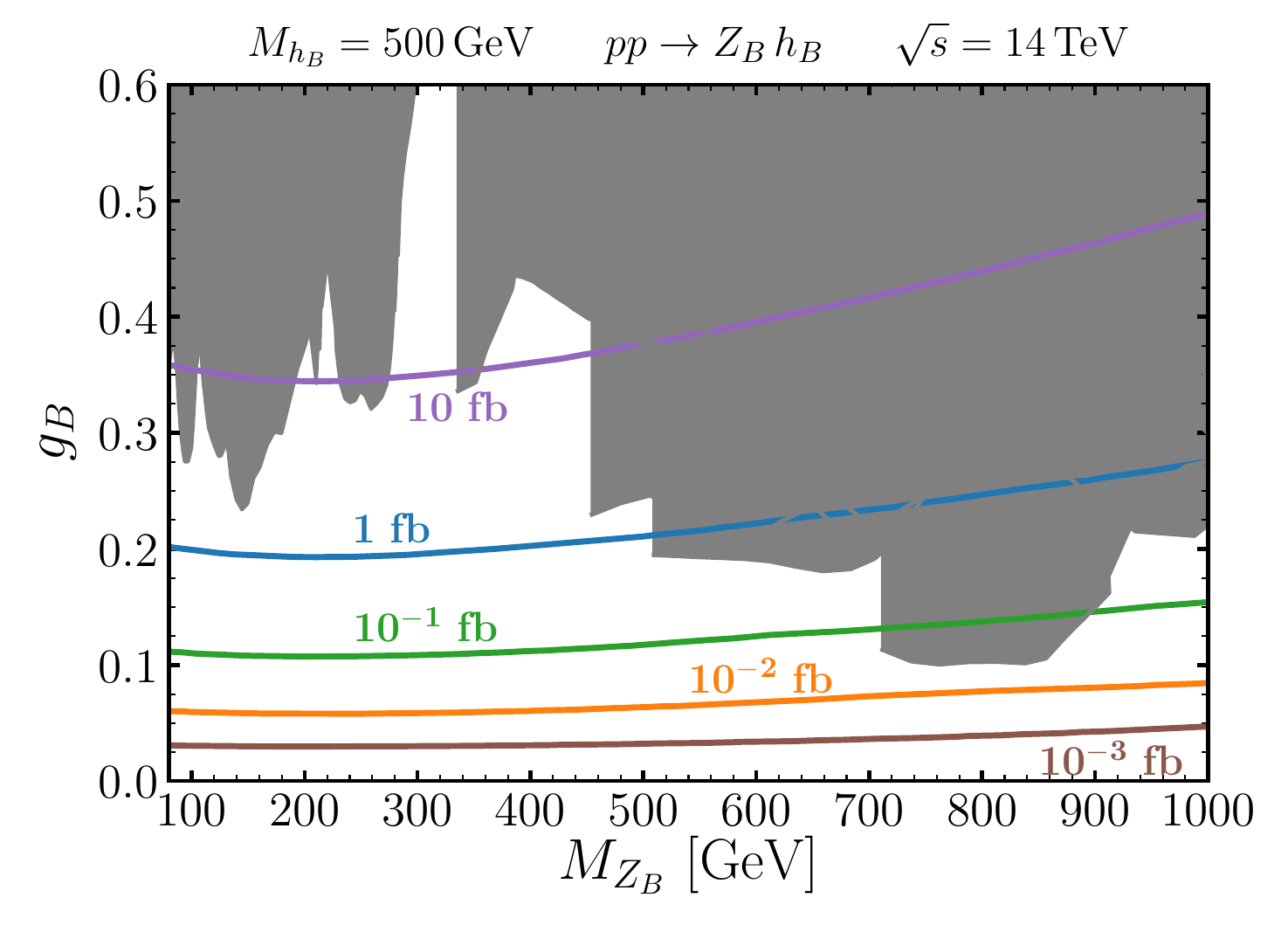}
\caption{Contour lines for the production cross-sections for the process $pp\to Z_B^* \to Z_B h_B$ at the LHC with center-of-mass energy of 14 TeV. These results have a very weak dependence on the mixing angle for $\sin \theta_B < 10^{-2}$. In the upper (lower) panel the scalar mixing angle is fixed to $\sin \theta_B = 10^{-3}$ (with identical results for $\sin \theta_B = 10^{-2}$). The region shaded in gray is excluded by direct searches for the $Z_B$ at the LHC.}
\label{ZBhB}
\end{figure}

The cross-sections are obtained using \texttt{MadGraph5aMC@NLO - v2.7.0}~\cite{Alwall:2014hca}, we cross-checked our results in a \texttt{Mathematica} notebook and the use of the MSTW2008~\cite{Martin:2009iq} set of parton distribution functions. In Fig.~\ref{ZBhB} we present the results for the cross-sections of the process $pp\to Z_B^* \to Z_B h_B$ in the $g_B$ vs $M_{Z_B}$ plane. For the upper (lower) panel we have fixed the mass of the second Higgs to 200 GeV (500 GeV), in agreement with the scenarios considered in Section~\ref{sec:DM}. The gray regions in the plot are excluded from direct searches of the $Z_B$ gauge boson by the ATLAS and CMS collaborations, for a discussion of these bounds see Ref.~\cite{FileviezPerez:2020mtk}. Since this cross-section scales as $\cos^2 \theta_B$, the results have a very mild dependence on the mixing angle for small values of $\theta_B$. 

\begin{figure}[h]
\centering
\includegraphics[width=0.65\linewidth]{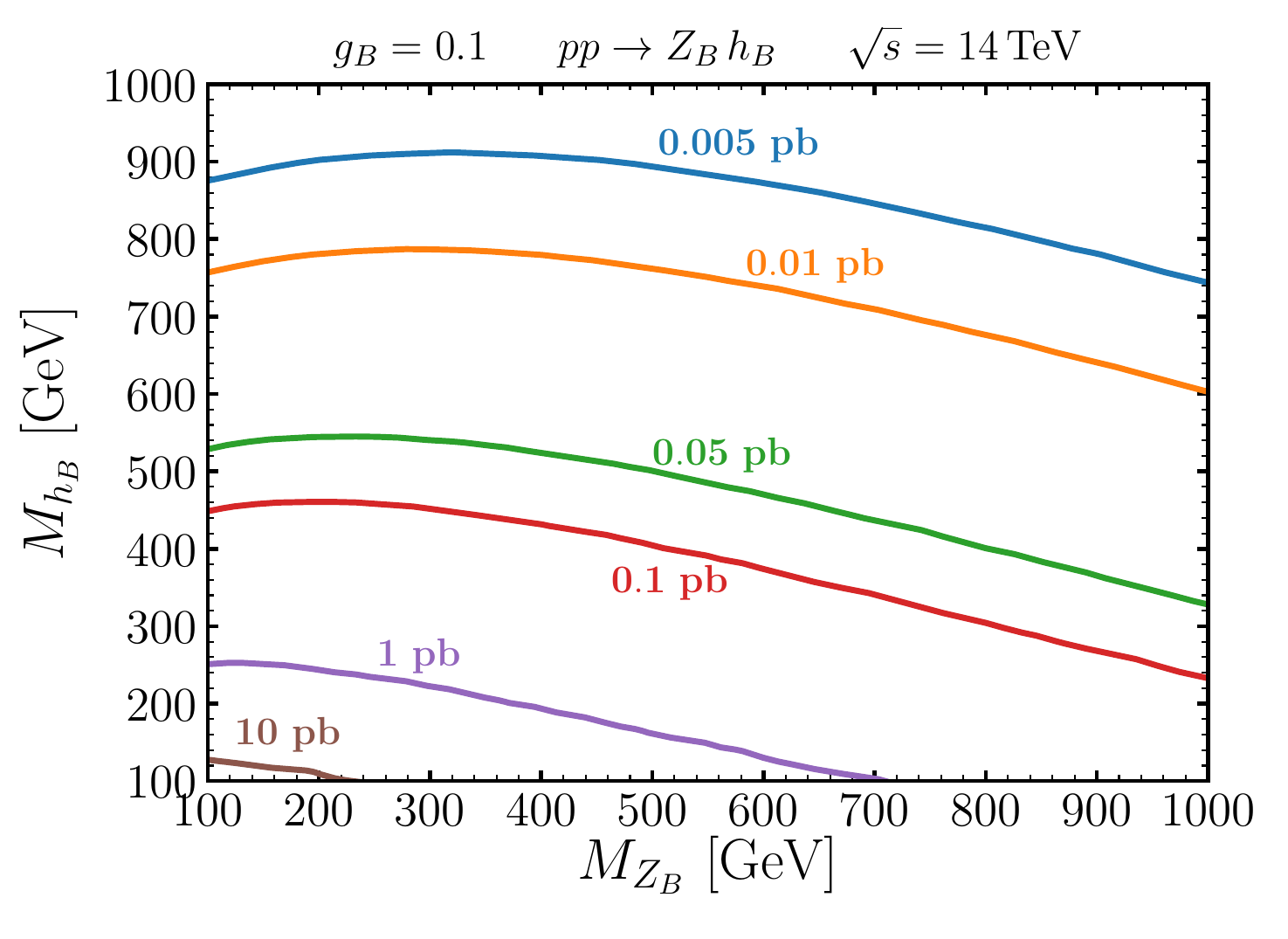}
\caption{Contour lines for the production cross-sections for the process $pp\to Z_B^* \to Z_B h_B$ at the LHC with center-of-mass energy of 14 TeV. We show the results in the $M_{Z_B}$ vs $M_{h_B}$ plane. These results have almost no dependence on the mixing angle for $\sin \theta_B < 10^{-2}$.}
\label{contours_xsection}
\end{figure}

In Fig.~\ref{contours_xsection} we present the result for the cross-section in the $M_{Z_B}$ vs $M_{h_B}$ plane. The gauge coupling has been fixed to $g_B=0.1$ in order to avoid the dijet constraints. As this plot shows for masses of the $Z_B$ around 100 GeV the production cross-section can be as large as 10 pb, while for masses close to a TeV the cross-section drops down to 0.005 pb.

\begin{figure}[h]
\centering
\includegraphics[width=0.495\linewidth]{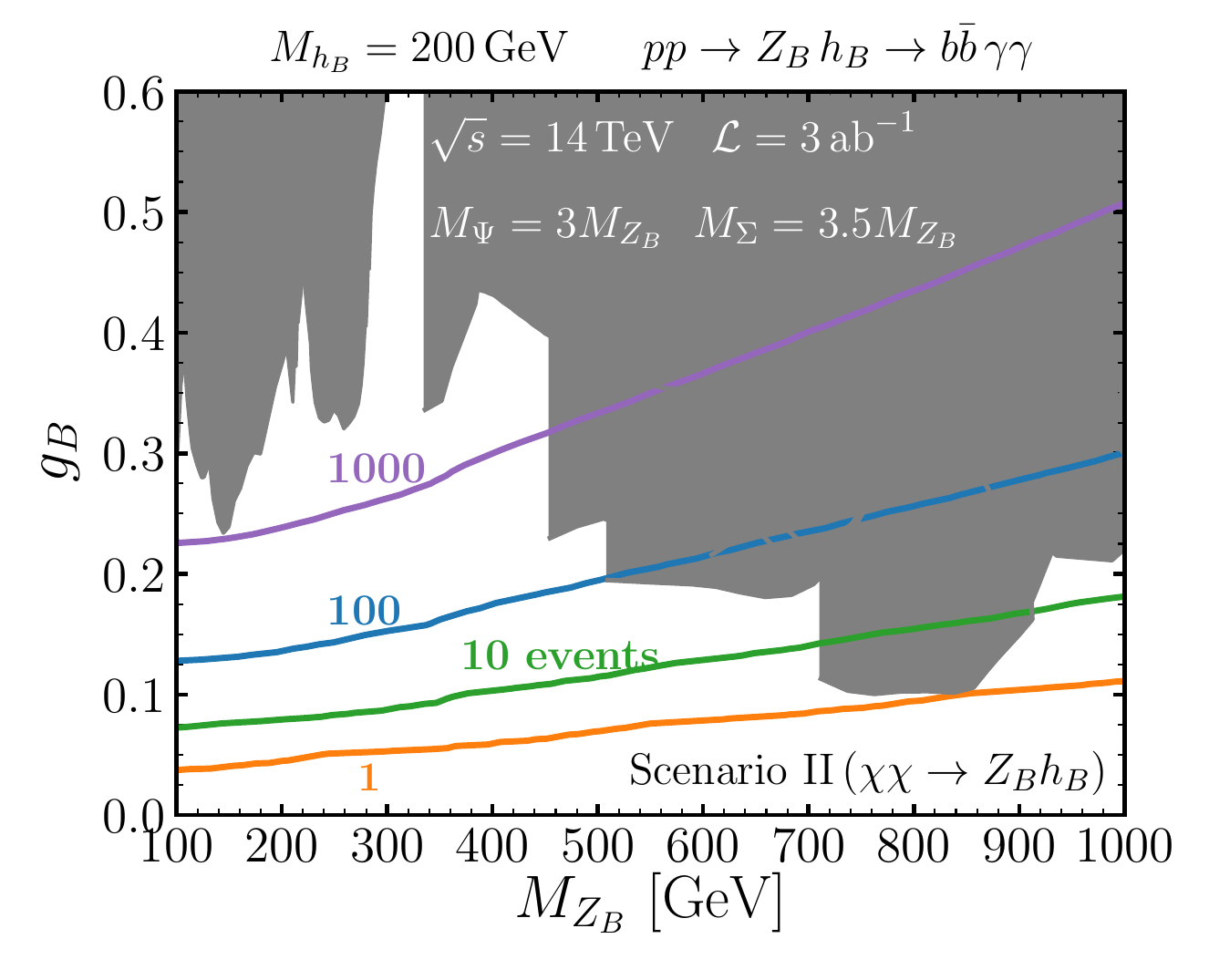}
\includegraphics[width=0.495\linewidth]{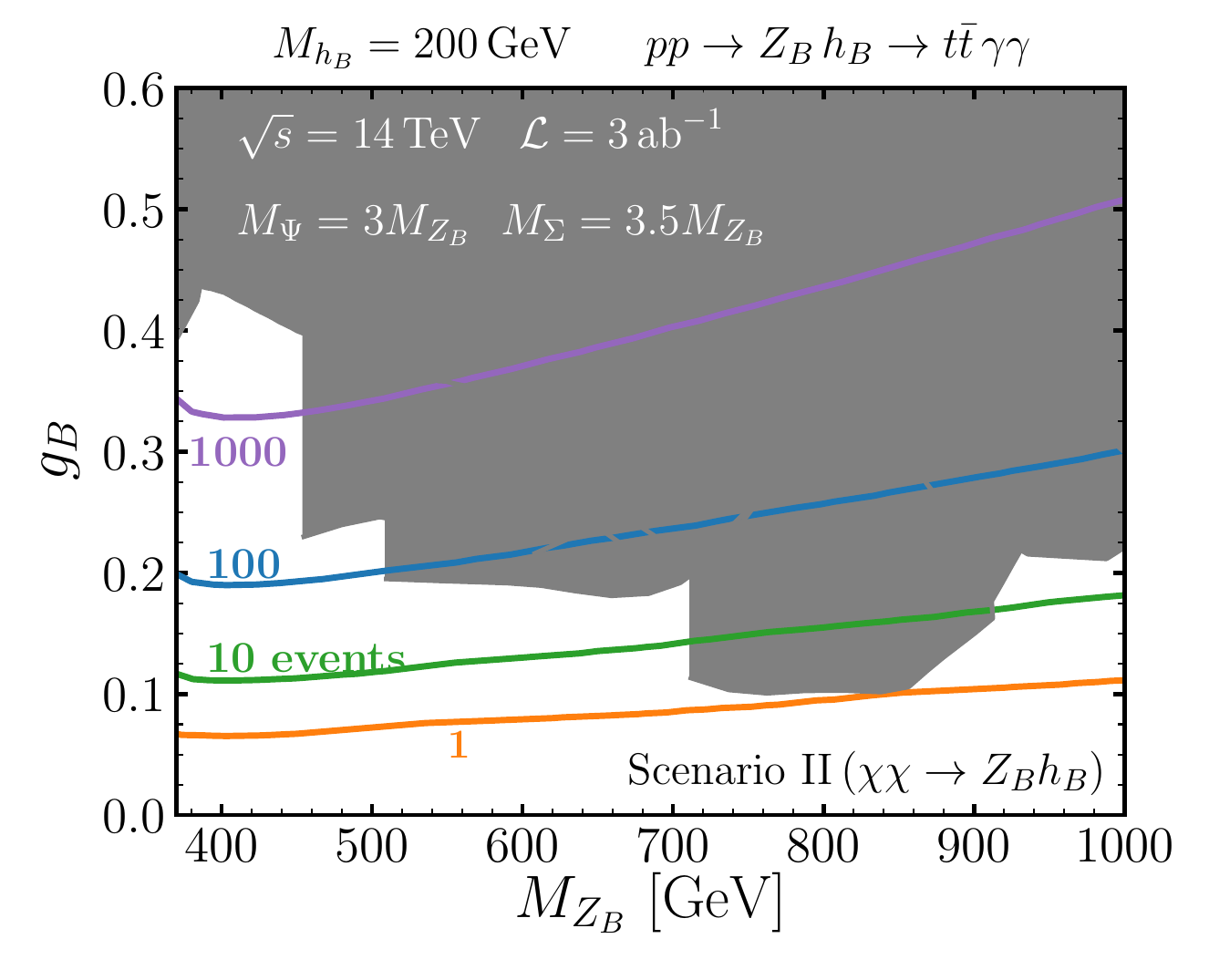}
\includegraphics[width=0.495\linewidth]{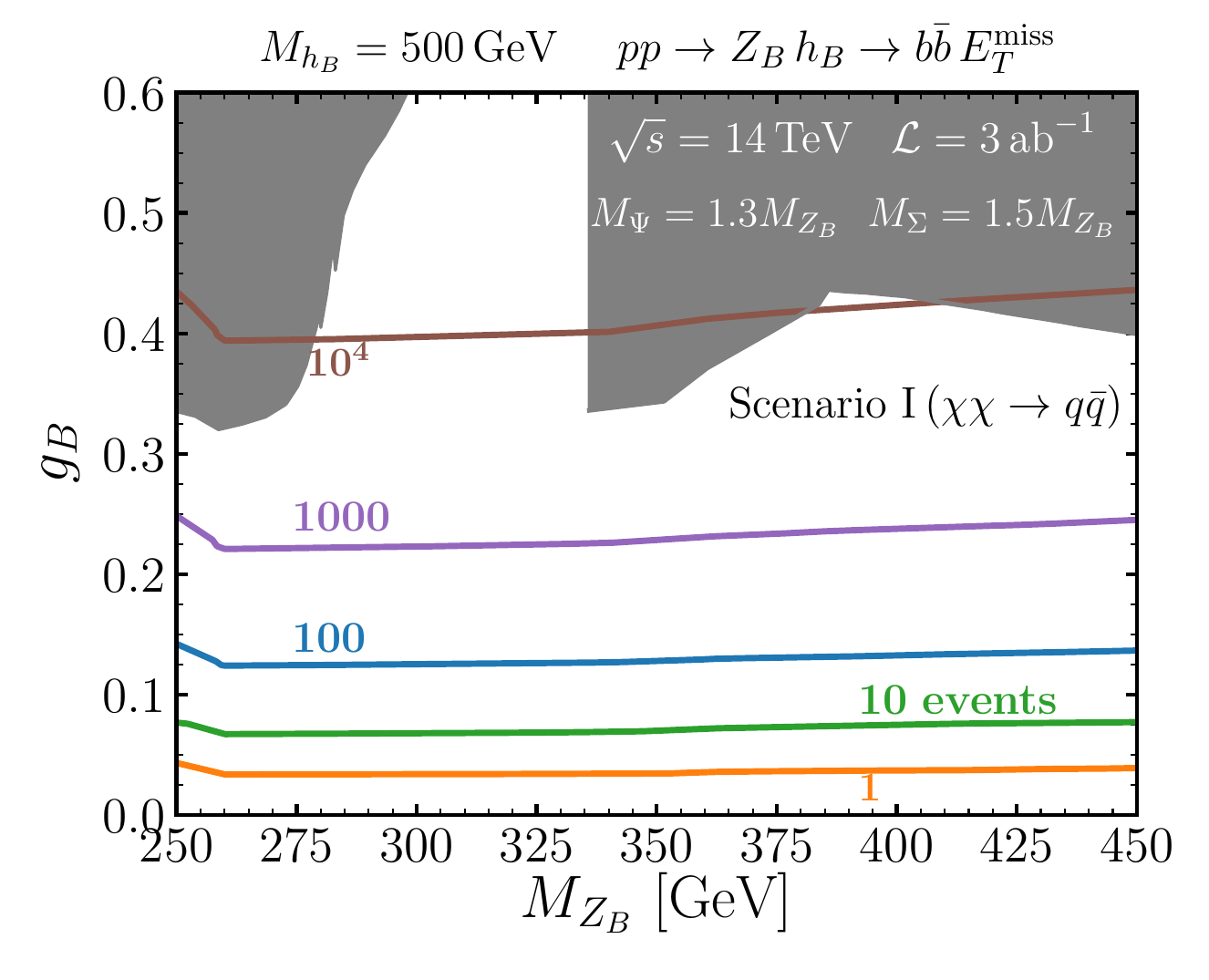}
\includegraphics[width=0.495\linewidth]{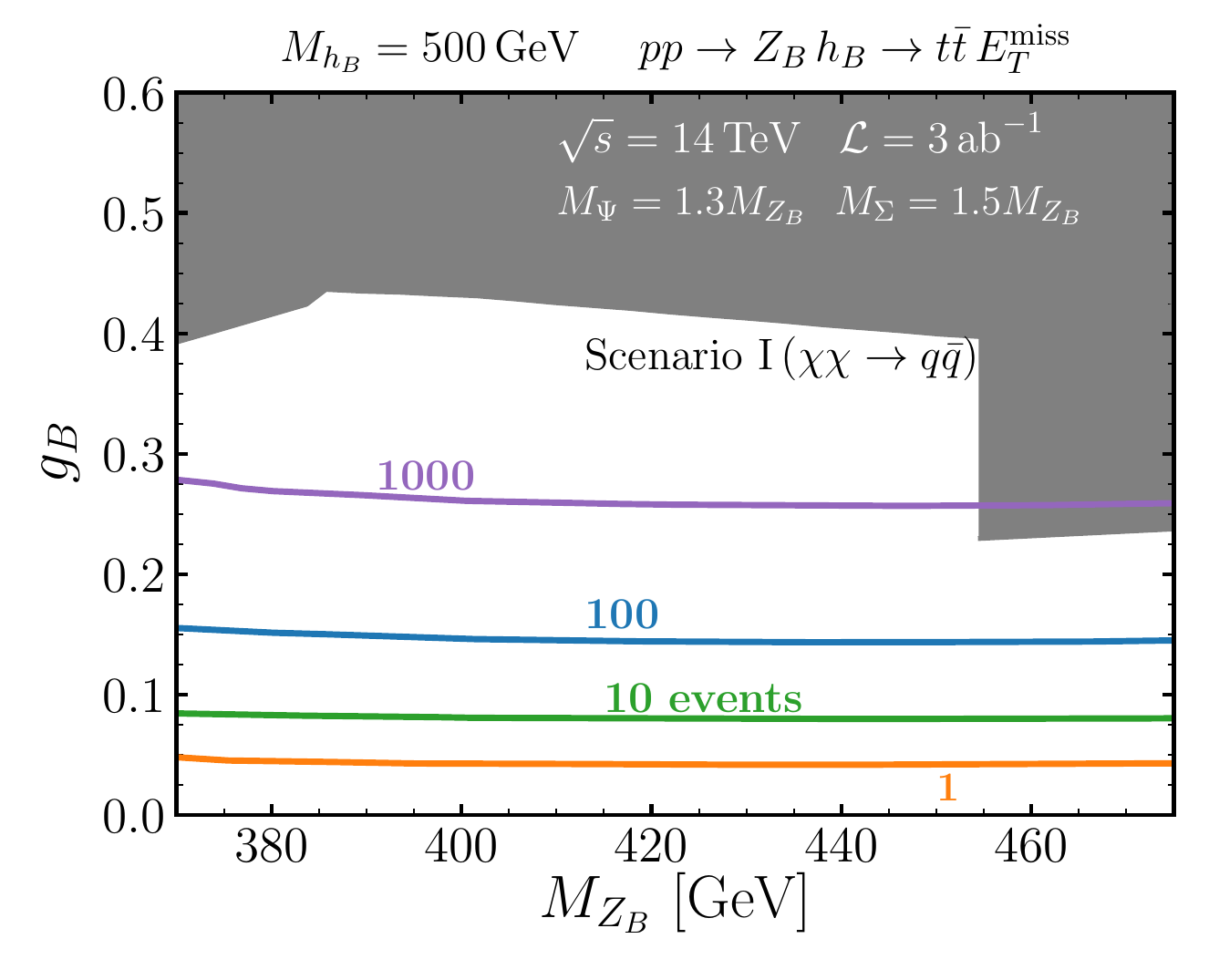}
\caption{Contour lines for the expected number of events at the LHC with center-of-mass energy of 14 TeV assuming an integrated luminosity of $\mathcal{L}=3000 \, {\rm fb}^{-1}$. We show the most relevant final states: For the upper panels we show $b\bar{b} \gamma\gamma$ and $t \bar{t} \gamma\gamma$ while for the lower panels we show  $b\bar{b} E_T^{\rm miss}$ and $t \bar{t} E_T^{\rm miss}$. In order to calculate the branching ratios of $h_B$ the masses of the new fermions have been scaled as the label shows on the plots. The gray region is excluded by the direct searches for a $Z_B$ at the LHC. The scalar mixing angle is fixed to $\sin \theta_B = 10^{-3}$ ($10^{-2}$) for the upper (lower) panels.}
\label{Nevents}
\end{figure}

The expected number of events at the LHC is given by,
\begin{equation}
N_{\rm events}(x\bar{x} y \bar{y})= {\mathcal{L}}  \times \sigma (p p \to Z_B^* \to Z_B  h_B)  \times { \rm BR}(Z_B\to x \bar{x}) \times {\rm BR}(h_B\to y\bar{y}).
\end{equation}
In Fig.~\ref{Nevents} we show our results for the expected number of events at the LHC with center-of-mass energy of 14 TeV and integrated luminosity of 3000 ${\rm fb}^{-1}$ for different combination of final states. The branching ratios for $Z_B$ and $h_B$ are computed at each point on the plots.

For the two plots in the upper panels of Fig.~\ref{Nevents} we consider Scenario II ($\chi \chi \to Z_B h_B$), and hence, we have that $2 M_\chi \gtrsim M_{h_B} + M_{Z_B}$. As we have discussed in Section~\ref{sec:DM}, in this context, if the Higgs is light enough such that the $Z_B Z_B$ channel remains closed, the branching ratio to two photons can be relevant. In order to compute the ${\rm BR}(h_B \to \gamma \gamma)$, we scale the masses of the anomaly-canceling fermions as $M_\Psi=3  \, M_{Z_B}$ and $M_\Sigma = 3.5 \, M_{Z_B}$. The dark matter mass remains a free parameter and can be fixed to the value that satisfies the correct dark matter relic density. In most of the parameter space in the plots we find that ${\rm BR}(h_B\to \gamma \gamma)\simeq 10^{-2}$.

For the two plots in the lower panels of Fig.~\ref{Nevents} we consider Scenario I ($\chi \chi \to q \bar{q} $) that is at the resonance $M_\chi \simeq M_{Z_B}/2$. Since we want the $h_B$ decay into missing energy to be the dominant one we only focus on the range for $M_{Z_B} = [M_{h_B}/2 , M_{h_B}]$. The masses of the anomaly-canceling fermions are scaled as $M_\Psi= 1.3 \, M_{Z_B}$ and $M_\Sigma = 1.5 \, M_{Z_B}$ in this context. In a large region of the parameter space considered we find that ${\rm BR}( h_B \to \chi \chi) \simeq 0.9$. For the left panel we consider the $b\bar{b}E_T^{\rm miss}$ final state for which the ATLAS collaboration~\cite{ATLAS:2019ivx} has a recent analysis, albeit for smaller Higgs masses that the ones we consider. The right panel shows our results for the $t\bar{t} E_T^{\rm miss}$ final state. In summary, in this theory it is possible to predict a large number of events in agreement with the collider bounds where the Baryonic Higgs has large branching ratios into two photons or into dark matter.

\subsection{The leptophobic gauge boson decays}
%

\begin{figure}[h]
\centering
\includegraphics[width=0.6\linewidth]{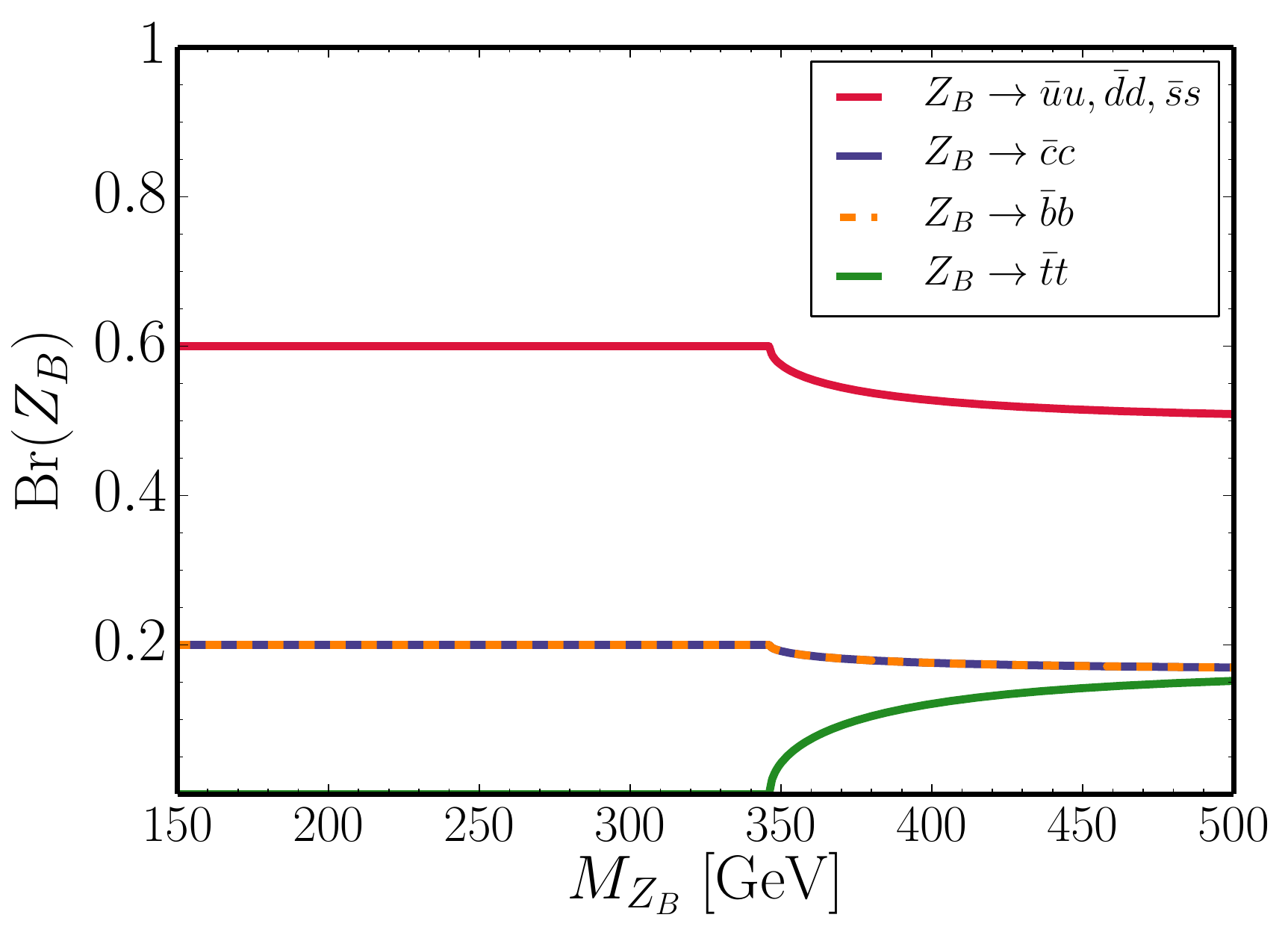}
\caption{Branching ratios of the $Z_B$ boson. In the parameter space where the relic density bound is satisfied, $\Omega h^2 \leq 0.12$, the $Z_B$ can only decay to quarks, and hence, the branching ratios into different quarks is completely determined as a function of the $Z_B$ mass.}
\label{BrZB}
\end{figure}

The new gauge boson in the theory $Z_B$ is coupled to the quarks and the anomaly-canceling fermions, and hence, $Z_B$ could decay into all of these particles including dark matter. However, in order to prevent overclosing the Universe we find that, as Fig.~\ref{fig:relic1} explicitly shows,
\begin{equation}
M_{Z_B} \lesssim 2 M_\chi.
\end{equation}
In consequence, the invisible decays of the $Z_B$ are kinematically forbidden and the gauge boson can decay only into the SM quarks. In Fig.~\ref{BrZB} we show the predictions of the branching ratios of the $Z_B$. We note also that these predictions do not depend on $g_B$ since this parameter factorizes out in the branching ratios.

We would like to emphasize that this simple prediction is a consequence of studying the correlation between the dark matter cosmological constraints and the predictions for collider physics.

\FloatBarrier

\section{Summary}
\label{sec:Summary}
In this work, we have discussed the possibility to search for the decays of a new Higgs in theories for dark matter where a Majorana dark matter is predicted from the cancellation of gauge anomalies. We have considered the minimal gauge theory for baryon number, in which four extra fermion representations are needed for anomaly cancellation, including a dark matter candidate. Taking into account these new fermions is crucial to compute the predictions for the decay of the new Higgs, since the loop-induced decays with the anomalons running in the loop can have large branching ratios in some regions of the parameter space.

We showed that there are three different scenarios  consistent with the relic density constraint; each one determined by the annihilation channel that gives the dominant contribution to the dark matter relic density.  In the scenario at the resonance $\chi \chi \to Z_B^* \to q \bar{q}$ it is possible for the Baryonic Higgs to decay mostly into a pair of dark matter particles so the invisible branching ratio can be very large $\text{BR}(h_B \to \chi \chi ) \simeq 90\%$.

In the second scenario, the annihilation channel $\chi \chi\to Z_B h_B$ dominates. This scenario is appealing since it does not rely on any resonance to achieve the correct dark matter relic abundance, and hence, a large region in the parameter space corresponds to this scenario. Furthermore, the branching ratio of the new Baryonic Higgs into photons can be much larger than the SM Higgs branching ratio into photons, i.e. $\text{BR}(h_B \to \gamma \gamma) \gtrsim 1\%$.

For the third scenario, the $\chi \chi \to Z_B Z_B$ is the dominant annihilation channel and the condition $M_{h_B}\lesssim 2 M_{Z_B}$ is satisfied, thus, the second Higgs can decays into a pair of the new gauge bosons. In fact, we find that the $h_B \to Z_B Z_B$ decay dominates, followed by the invisible decay $h_B \to \chi \chi$ which, if kinematically open, usually has a branching ratio $\text{BR}(h_B \to \chi \chi ) \simeq 10\%$. 

We also demonstrated that the dark matter direct detection experimental bounds set a strong limit on the mixing between the SM Higgs and the Baryonic Higgs. Consequently, all the SM-like Higgs production mechanisms are typically suppressed. Motivated by this, we studied the cross section for the associated production $p p \to Z_B^* \to h_B Z_B$ 
and showed that it can be large in agreement with all experimental constraints. Therefore, it is possible to expect a large number of events for the exotic signatures $ t\bar{t} E_T^{miss}$ and $b\bar{b} E_T^{miss}$ in the scenarios where the invisible branching ratio can be large, and $t\bar{t} \gamma \gamma$ and $b\bar{b} \gamma \gamma$ in the  scenario where the Baryonic Higgs can decay into two photons with a large branching ratio. These results further motivate new studies in the search for a new Higgs that decays into dark matter at ATLAS and CMS. In summary, our results show the importance of the correlation between the cosmological constraints 
and the predictions for Higgs decays in a theory predicting the existence of dark matter from anomaly cancellation.

\vspace{0.7cm}
{\textit{Acknowledgments}}: The work of P.F.P. has been supported
by the U.S. Department of Energy, Office of Science, Office of
High Energy Physics, under Award Number de-sc0020443. This material is based upon work supported by the U.S. Department of Energy, Office of Science, Office of High Energy Physics, under Award Number DE-SC0011632. C.M. thanks the support provided by the Walter Burke Institute for Theoretical Physics. We would like to thank E. Golias for several discussions.

\appendix

\section{Feynman Rules}
\label{sec:appFR}
%
In this appendix we list the Feynman rules that have been used in this work. In the following expressions, as well as in Appendix~\ref{sec:appDecays}, we define the dark matter field and the Majorana anomalons as:
\begin{equation}
\chi = \chi^0_L + (\chi^0_L)^C , \quad \quad \text{ and }\quad \quad \Sigma^0 = \Sigma^0_L + (\Sigma^0_L)^C,
\end{equation}
while the Dirac anomalons are given by
\begin{equation}
\Psi^+ = (\Psi_1^+)_L + (\Psi_2^+)_R, \quad \quad \text{ and }\quad \quad \Sigma^+ = \Sigma^+_L + (\Sigma_L^-)^C.
\end{equation}
 Taking the above definitions into account and working in the context where we neglect the Yukawa couplings between the new fermions and the SM Higgs, $y_i$ with $i=1,...,4$ from Eq.~\eqref{eq:LB}, the simplified Feynman rules read as:

 \begin{equation*}
\begin{aligned}[c]
&\imineq{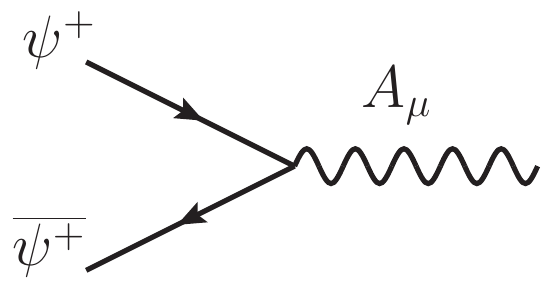}{12}:\quad  -ie\gamma^\mu,\\[2ex]
&\imineq{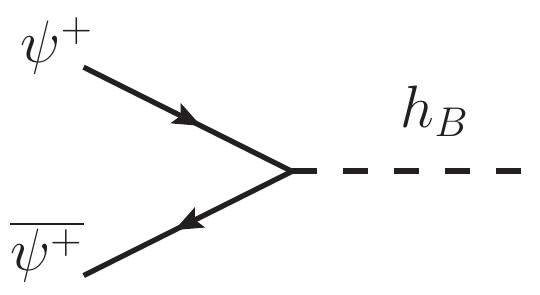}{12}: \quad - \frac{i}{\sqrt{2}} \cos \theta_B y_\Psi ,
\end{aligned}
\qquad \qquad 
\begin{aligned}[c]
&\imineq{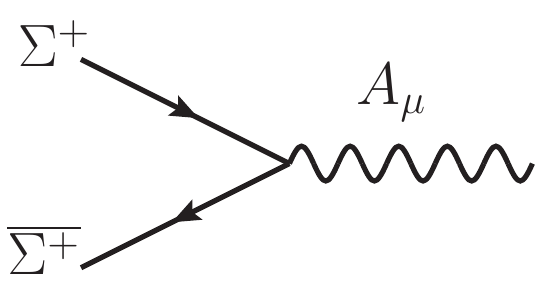}{12}: \quad -ie\gamma^\mu, \nonumber \\[2ex]
&\imineq{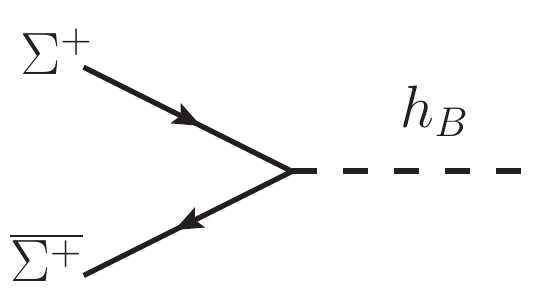}{12}: \quad -i \sqrt{2} \cos \theta_B  y_\Sigma,  \nonumber \\
\end{aligned}
\end{equation*}

\begin{equation*}
\begin{aligned}[c]
&\imineq{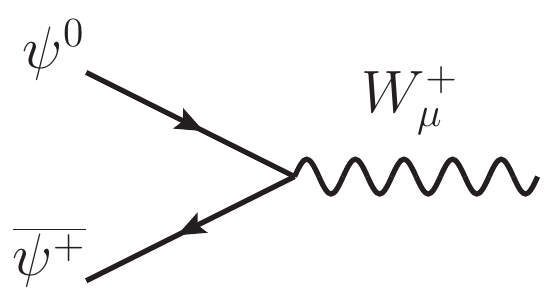}{12}: \quad - \frac{ig} {\sqrt{2}} \gamma^\mu,\\[2ex]
&\imineq{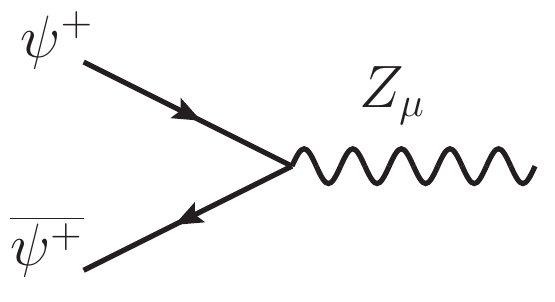}{12}: \quad - i g \frac{s_W}{t_{2W}} \gamma^\mu,\\[2ex]
&\imineq{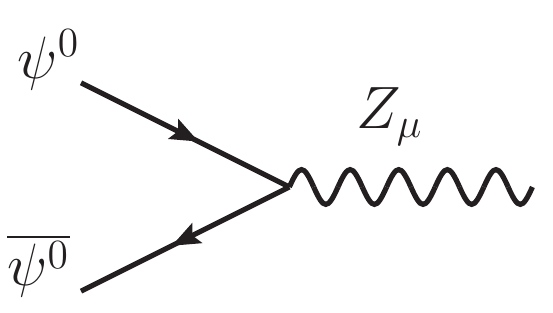}{12}: \quad \frac{ig}{2c_W}\gamma^\mu, 
\end{aligned}
\qquad \qquad
\begin{aligned}[c]
&\imineq{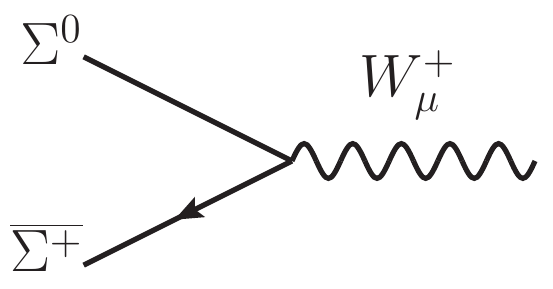}{12}: \quad  ig \gamma^\mu,   \nonumber \\[2ex]
&\imineq{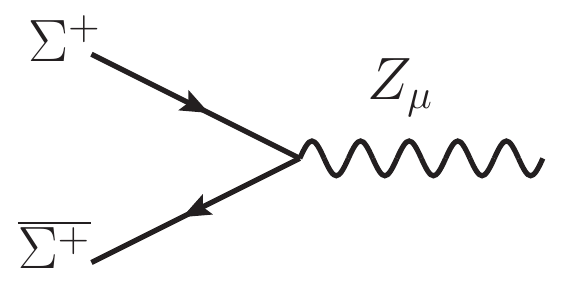}{12}: \quad - i g \frac{s_W}{t_{W}} \gamma^\mu, \nonumber\\
&\imineq{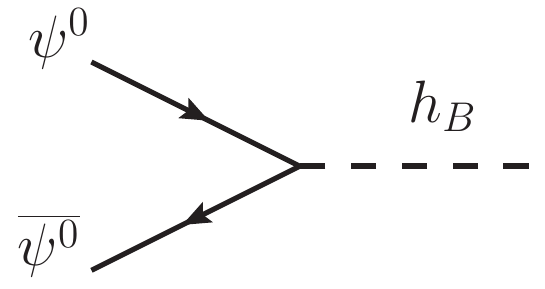}{12}: \quad - \frac{i}{\sqrt{2}} \cos \theta_B y_\Psi.  \nonumber
\end{aligned}
\end{equation*}



%
%


%
%
\section{Higgs Decays}
\label{sec:appDecays}
%
In this Appendix, we present the Higgs decays including the loop-induced channels and the full analytic expressions. To compute the loop functions we make use of the Package-X~\cite{Patel:2015tea} Mathematica package. In the Feynman diagrams below, the orange dot symbolizes that the vertex can only occur by the mixing of the Baryonic Higgs with the SM Higgs.
\begin{itemize}[leftmargin=3mm]
%
\item ${\boldmath h_B \to \gamma \gamma}$:
\begin{eqnarray}
\imineq{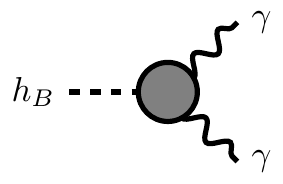}{11} &=& \imineq{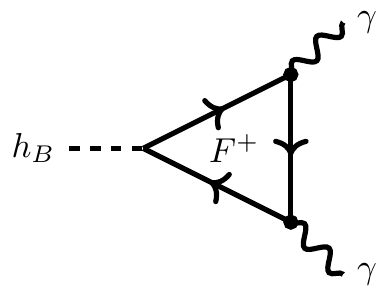}{17}+\imineq{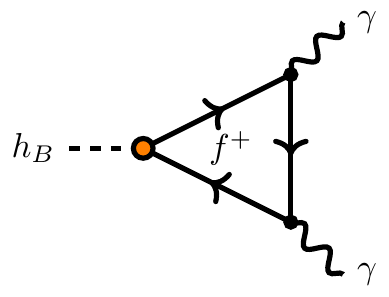}{17} + \imineq{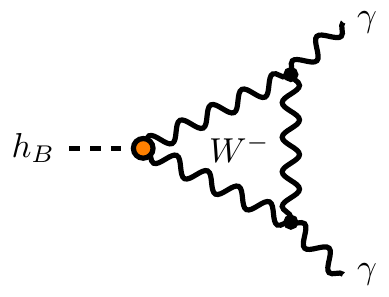}{17} \nonumber  \\
\Gamma(h_B \to \gamma \gamma) &=&  \frac{\alpha^2}{64  \pi^3  M_{h_B}^5} \left |  \cos \theta_B \sum_{F^+}  \frac{3 g_B M_{F^+}^2}{M_{Z_B}} F_{F^+} + \frac{\sin \theta_B}{v_0} \left( \sum_{f^+} N_c^f Q_f^2 m_{f^+}^2 F_{f^+} -   F_W \right) \right |^2.  \quad \quad \quad \nonumber 
\end{eqnarray}
%
\item $\boldmath h_B \to Z_B \gamma$:
\begin{eqnarray}
\imineq{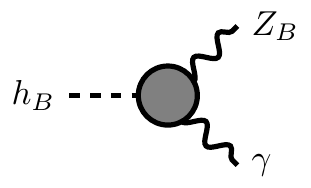}{11} &=& \underbrace{\imineq{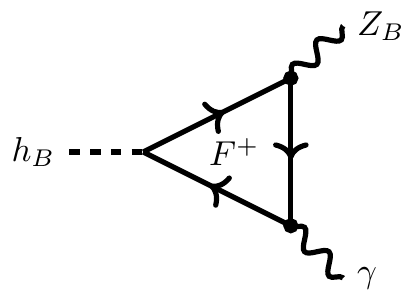}{17}}_{=0} + \imineq{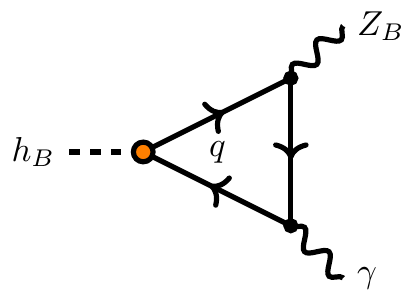}{17} \nonumber \\
\Gamma(h_B \to Z_B \gamma ) &=& \sin^2 \theta_B \, g_B^2 \, \frac{\alpha}{288 \, \pi^4}  \, \frac{M_{h_B}^2-M_{Z_B}^2}{v_0^2 M_{h_B}^3}\left | \sum_q  Q_q m_q^2  A_q \right |^2. \nonumber
\end{eqnarray}
\end{itemize}

\begin{itemize}[leftmargin=3mm]
\item  $h_B \to Z \gamma$:
\begin{eqnarray}
 \!\!\!\!  \imineq{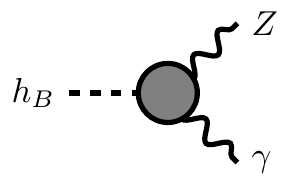}{11} \!\!\!\!\! &=& \imineq{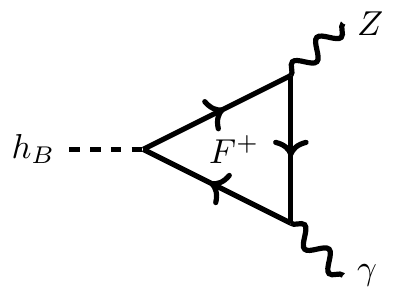}{17} \!\!\!\!\! + \imineq{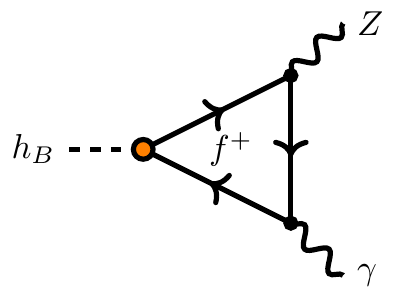}{17} \!\!\!\!\!+  \imineq{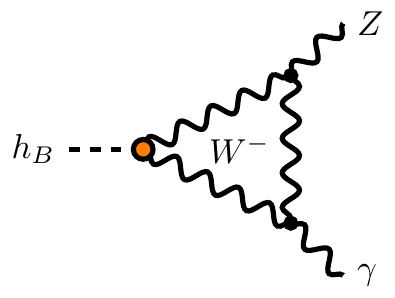}{17} \quad \quad \quad \quad \quad \quad  \quad  \quad \quad \quad \quad   \nonumber \\
  \!\!\!\!  \!\!\!\! \Gamma(h_B \to Z \gamma ) &=& \frac{\alpha^2}{8\pi^3} \frac{M_{h_B}^2 \! - \! M_{Z}^2}{M_{h_B}^3}  \left |  \frac{\sin \theta_B}{v_0 \, s2\theta_W} \left( \sum_{u}  m_u^2 \left(1 \! - \! \frac{8}{3} \sin^2 \theta_W\right)  A_u + \sum_{d} \frac{m_d^2}{2} \left(1 \! - \! \frac{4}{3}\sin^2 \theta_W \right) A_d   \right. \right. \quad \quad \quad \nonumber \\
& + & \left. \sum_{e} \frac{m_e^2}{2} \left(1 \! - \! 4 \sin^2 \theta_W \right) A_e - \frac{\cos^2\theta_W}{2} A_W \right)
+ \left. \frac{3 \, g_B\cos \theta_B}{M_{Z_B}} \left( \frac{M_{\Sigma^+}^2}{t \theta_W}A_{\Sigma^+} +  \frac{M_{\Psi^+}^2}{t2\theta_W}A_{\Psi^+} \right) \right |^2. \nonumber
\end{eqnarray}  
%
 \item  $h_B \to g g$: 
  \begin{eqnarray}
  \centering
  && \imineq{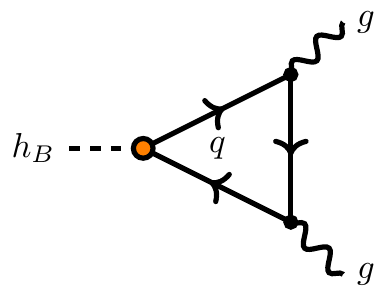}{15} \qquad  \Gamma(h_B \to g g) =  \sin^2 \theta_B  \frac{ \alpha_s^2 M_{h_B}^3}{128 \, v_0^2 \, \pi^3} \left | \sum_q \frac{2 m_q^2}{M_{h_B}^4} F_q \right |^2. \nonumber
  \end{eqnarray} 
  %
 \item $h_B \to Z_B Z_B$:
  \begin{eqnarray}
  \centering
&& \imineq{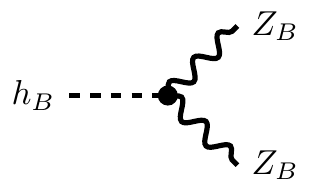}{12}  \qquad \Gamma(h_B \to Z_B Z_B) = \frac{9 g_B^2}{32\pi} \cos^2\theta_B \sqrt{1-\frac{4M_{Z_B}^2}{M_{h_B}^2}} \, \frac{M_{h_B}^4-4M_{h_B}^2M_{Z_B}^2+12M_{Z_B}^4}{M_{h_B} M_{Z_B}^2}. \quad \quad \quad \nonumber
\end{eqnarray}
%
\item $h_B \to WW$:
%
\begin{eqnarray}
 \imineq{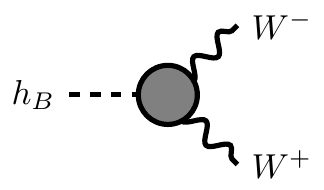}{12} &=& \imineq{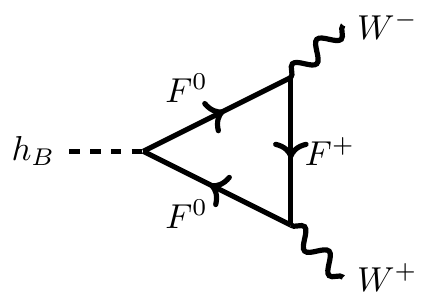}{15}+\imineq{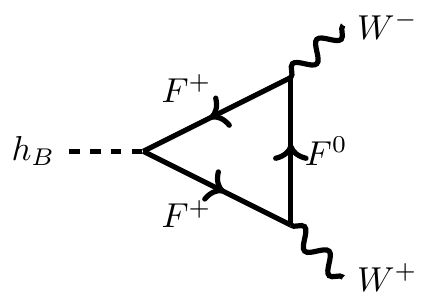}{15}+ \imineq{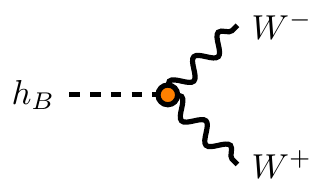}{12} \nonumber \\
\Gamma (h_B \to WW) &=& \frac{\sqrt{M_{h_B}^2 - 4M_W^2}}{16\pi M_{h_B}^2 M_W^4} \left (\cos^2 \theta_B \frac{9g_B^2}{M_{Z_B}^2}\left | \sum_F g_{WF}^2 B_F[W] \right |^2 +2(M_{h_B}^2-2M_W^2) \times \right. \nonumber \\
&&\text{Re} \left \{ \cos \theta_B \frac{3 g_B}{M_{Z_B}} \sum_F g_{WF}^2 B_F[W] \left ( \cos \theta_B \frac{3g_B}{M_{Z_B}} \sum_F g_{WF}^2 C_F^*[W] + \frac{\sin \theta_B}{v_0} M_W^2 \right) \right \}  \nonumber \\
&& \left.+ \left |  \cos \theta_B \frac{3g_B}{M_{Z_B}} \sum_F g_{WF}^2 C_F[W] +  \frac{\sin \theta_B}{v_0} M_W^2 \right|^2(M_{h_B}^4 - 4 M_{h_B}^2 M_W^2 + 12 M_W^4) \right). \nonumber
\end{eqnarray}
\end{itemize}

\begin{itemize}[leftmargin=3mm]
\item $h_B \to ZZ$:
%
\begin{eqnarray}
\imineq{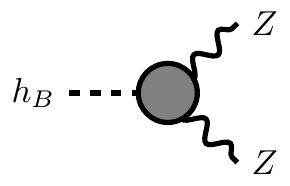}{12}
&=&
\imineq{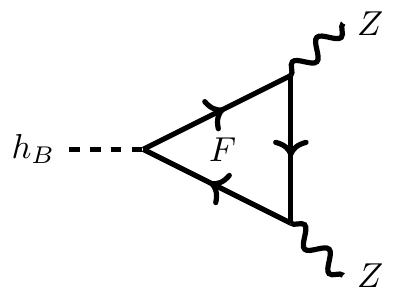}{15}
+
\imineq{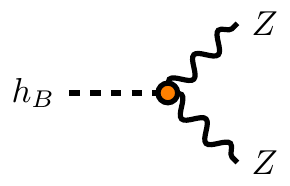}{12} \nonumber
\\
\Gamma (h_B \to ZZ) &=& \frac{\sqrt{M_{h_B}^2 - 4M_Z^2}}{128\pi M_{h_B}^2 M_Z^4} \left ( \cos^2\theta_B \frac{9g_B^2}{M_{Z_B}^2}\left |\sum_F g_{ZF}^2 B_F[Z] \right |^2 + 2 (M_{h_B}^2-2M_Z^2) \times \right. \nonumber  \\
&&\text{Re} \left \{ \cos \theta_B \frac{3 g_B}{M_{Z_B}} \sum_F g_{ZF}^2  B_F[Z] \left ( \cos \theta_B \frac{3g_B}{M_{Z_B}} \sum_F g_{ZF}^2 C_F^*[Z] + 2 \frac{\sin \theta_B}{v_0} M_Z^2 \right) \right \} \nonumber \\
&& \left.+ \left |\cos \theta_B \frac{3g_B}{M_{Z_B}} \sum_F g_{ZF}^2 C_F[Z]  + 2  \frac{\sin \theta_B}{v_0} M_Z^2 \right|^2(M_{h_B}^4 - 4 M_{h_B}^2 M_Z^2 + 12 M_Z^4) \right). \nonumber
\end{eqnarray}
%
\item $h_B \to Z Z_B$:
%
\begin{eqnarray}
\imineq{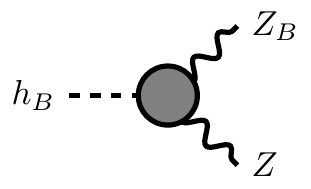}{12}
&=&
\underbrace{
\imineq{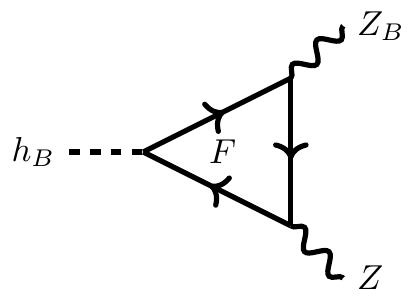}{15}}_{= \, 0}
+
\imineq{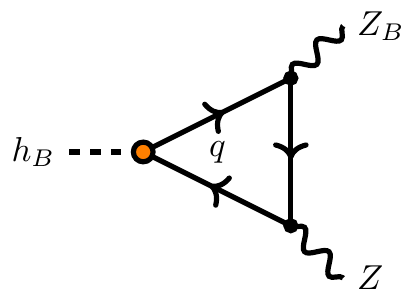}{15} \nonumber \\
\Gamma(h_B \to Z Z_B)&=& \sin^2\theta_B  \, g_B^2 \,  \alpha \, \frac{ \lambda^{1/2}(M_{h_B}^2,M_Z^2,M_{Z_B}^2)}{64  \, v_0^2 \, \sin^22\theta_W M_{h_B}^3 M_Z^2 M_{Z_B}^2} \nonumber \\
&& \left( \left| \tilde \Gamma_1^{ZZ_B} \right |^2  - 2 \, \text{Re} \left \{\tilde  \Gamma_1^{ZZ_B} ( \tilde \Gamma_2^{ZZ_B})^* \right \} (M_{h_B}^2-M_Z^2-M_{Z_B}^2) \right. \nonumber \\
&& \left. + \left| \tilde \Gamma_2^{ZZ_B} \right |^2 (M_{h_B}^4 - 2 M_{h_B}^2 (M_Z^2 + M_{Z_B}^2) + M_Z^4 + 10 \, M_Z^2 M_{Z_B}^2 + M_{Z_B}^4 ) \right), \nonumber
\end{eqnarray}
where
%
\begin{equation*}
\tilde \Gamma_1^{ZZ_B} \equiv    \sum_q v_Z^q m_q^2 D_q, \quad \text{ and }  \quad \tilde \Gamma_2^{ZZ_B} \equiv   \sum_q v_Z^q  m_q^2 E_q .
\end{equation*}
\end{itemize}

Below we write explicitly the loop functions as a function of the fermion mass $m_f$ or gauge boson mass $M_V$ running inside the loop and the Passarino-Veltman functions defined in Package-X:
\begin{itemize}[leftmargin=3mm]
\item Loop functions entering in the $h_B$ decay to two massless gauge fields:
\begin{eqnarray*}
 F_f &=&  4 M_{h_B}^2 + (4 M_f^2 - M_{h_B}^2) \log^2 \left[1-\frac{M_{h_B}^2}{2M_f^2}\left( 1+\sqrt{1 - \frac{4M_f^2}{M_{h_B}^2 \, }} \right) \right],
 \end{eqnarray*}
 \begin{eqnarray*}
 F_W &=& M_{h_B}^4 + 6 M_{h_B}^2 M_W^2 - 3 (M_{h_B}^2 - 2 M_W^2)M_W^2 \log^2 \left[1- \frac{M_{h_B}^2}{2M_W^2}\left( 1 + \sqrt{1 -\frac{ 4M_W^2}{M_{h_B}^2}} \, \right)\right].
 \end{eqnarray*}
 \item Loop functions entering in the $h_B$ decay to a massive and a massless gauge fields:
 \begin{eqnarray*}
 A_f &=& (4m_f^2-M_{h_B}^2+M_{V}^2)C_0[0,M_{h_B}^2,M_{V}^2;m_f] + 2\left(\frac{M_{V}^2(\Lambda[M_{h_B}^2]-\Lambda[M_{V}^2])}{M_{h_B}^2- M^2_{V}}+1\right),\\
 A_W &=& 2 \, M_W^2 \, C_0[0,M_{h_B}^2,M_{V}^2;M_W] \left( (M_{h_B}^2-2M_W^2)(\tan^2\theta_W-5) - 2 M_V^2 (\tan^2\theta_W-3)\right)\\
 && +\left( \frac{M_V^2}{M_{h_B}^2-M_V^2} (\Lambda[M_{h_B}^2]-\Lambda[M_V^2]) +1\right) \left(M_{h_B}^2(1-\tan^2\theta_W)+2M_W^2 (5-\tan^2\theta_W)\right).
 \end{eqnarray*}
 \item Loop functions entering in the $h_B$ decay to a two equal massive gauge bosons:
 \begin{eqnarray*}
 B_F[V] &=&\frac{1}{16 \pi^2} \frac{4M_F^2}{M_{h_B}^2-4M_V^2} \times \\ 
 && \left[ (M_{h_B}^2-2M_V^2) \left( M_{h_B}^4 - 4 M_F^2 (M_{h_B}^2-4M_V^2) - 6 M_{h_B}^2M_V^2-4M_V^4\right) C_0[M_{h_B}^2,M_V^2,M_V^2;M_F] \right. \\
&& \left. + \, 4 M_V^2 (M_{h_B}^2 + 2M_V^2)(\Lambda[M_V^2]-\Lambda[M_{h_B}^2])-2(M_{h_B}^4-6M_{h_B}^2M_V^2+8M_V^4)\right),\\
C_F[V] &=& \frac{1}{16 \pi^2} \frac{4M_F^2}{M_{h_B}^2-4M_V^2} \left[ 2 (M_{h_B}^2 + 2M_V^2 (\Lambda[M_{h_B}^2]-\Lambda[M_V^2]) -4M_V^2)\right. \\
&& \left. - (M_{h_B}^4 - 6 M_{h_B}^2 M_V^2 - 4 M_F^2 (M_{h_B}^2 - 4 M_V^2) + 4 M_V^4)C_0 [M_{h_B}^2,M_V^2,M_V^2;M_F]\right].
\end{eqnarray*}
\item Loop functions entering in the decay to two different massive gauge boson:
\begin{eqnarray*}
D_q &=& \frac{(M_{h_B}^2-(M_Z-M_{Z_B})^2)(M_{h_B}^2-(M_Z+M_{Z_B})^2)}{ \pi^2 (4 M_Z^2 M_{Z_B}^2-(M_{h_B}^2-M_Z^2 -M_{Z_B}^2)^2)^2}  \times \\
&& \left[ ( (M_{h_B}^2-M_Z^2-M_{Z_B}^2) \left(M_{h_B}^6 - 4 m_q^2(M_{h_B}^4 - 2M_{h_B}^2 (M_Z^2 + M_{Z_B}^2)   + (M_Z^2-M_{Z_B}^2)^2 \right) \right. \\ 
&& - 3 M_{h_B}^4 (M_Z^2 + M_{Z_B}^2) + M_{h_B}^2 (3 M_Z^4 - 10 M_Z^2 M_{Z_B}^2 + 3 M_{Z_B}^4) \\
&&- (M_Z^2-M_{Z_B}^2)^2(M_Z^2 + M_{Z_B}^2) ) C_0[M_{h_B}^2,M_Z^2,M_{Z_B}^2;m_q] /2 \\
&& + M_Z^2 (4 M_{Z_B}^2(M_{h_B}^2+M_Z^2) + (M_{h_B}^2-M_Z^2)^2-5M_{Z_B}^4 )\Lambda[M_Z^2]\\
&& -\left (M_{h_B}^4(M_Z^2+M_{Z_B}^2)- 2 M_{h_B}^2 (M_Z^4 - 4 M_Z^2 M_{Z_B}^2 + M_{Z_B}^4) + (M_Z^2-M_{Z_B}^2)^2(M_Z^2+M_{Z_B}^2 \right ) \Lambda[M_{h_B}^2] \\
&& + M_{Z_B}^2 \left( M_{h_B}^4 + M_{h_B}^2 ( 4 M_Z^2 - 2 M_{Z_B}^2 )-5 M_Z^4 + 4 M_Z^2 M_{Z_B}^2 + M_{Z_B}^4 \right ) \Lambda[M_{Z_B}^2] \\
&& \left. +  (M_{h_B}^2-M_Z^2 - M_{Z_B}^2) \left ( 4 M_Z^2 M_{Z_B}^2 -(M_{h_B}^2-M_Z^2-M_{Z_B}^2)^2 \right) \right],\\
E_q &=&  \displaystyle  \left[  \phantom{\frac{A}{B}} \!\!\!\!\!\!\!\! \left(4 M_{h_B}^2 M_Z^2 M_{Z_B}^2 + (4 m_q^2 - M_{h_B}^2 + M_Z^2 + M_{Z_B}^2 ) \lambda[M_{h_B}^2, M_Z^2,M_{Z_B}^2] \right )C_0[M_{h_B}^2,M_Z^2,M_{Z_B}^2;m_q] \right. \\
&&- 2 M_Z^2 (M_{h_B}^2-M_Z^2+M_{Z_B}^2)\Lambda [M_Z^2] + 2 \left ( M_{h_B}^2 (M_Z^2+M_{Z_B}^2)-(M_Z^2-M_{Z_B}^2)^2\right) \Lambda[M_{h_B}] \\
&& \left. \phantom{\frac{A}{B}}\!\!\!\!\!\!\!\!\! - 2 M_{Z_B}^2 (M_{h_B}^2 + M_Z^2 - M_{Z_B}^2)\Lambda[M_{Z_B}^2] + 2 \lambda[M_{h_B}^2,M_Z^2,M_{Z_B}^2] \right] \big / \left( 4\pi^2 \lambda[M_{h_B}^2,M_Z^2,M_{Z_B}^2] \right),
 \end{eqnarray*}
where $ v_Z^q = T_3^q(1-4|Q_q|\sin^2\theta_W)$ is the vector coupling of the quarks with the $Z$ boson and $T_3^q$ is the weak isospin, and the K\"allen function is defined as follows:
 \begin{equation}
 \lambda[M_{h_B}^2, M_Z^2,M_{Z_B}^2] = M_{h_B}^4-2 M_{h_B}^2 M_Z^2-2 M_{h_B}^2 M_{Z_B}^2+M_Z^4-2 M_Z^2 M_{Z_B}^2+M_{Z_B}^4.
 \end{equation}
 \end{itemize}
Finally we list the baryonic Higgs decays to SM fermions, to the anomalons and to a couple of SM Higgs bosons:
 \begin{eqnarray*}
\Gamma(h_B \to \bar f f) &=&  N_c^f  \, \sin^2 \theta_B \frac{ m_f^2}{8\pi \, v^2}M_{h_B} \left(1-4 \frac{m_f^2}{M_{h_B}^2}\right)^{3/2},\\[1.2ex]
\Gamma(h_B \to \bar \Psi^0 \Psi^0,\bar \Psi^+ \Psi^+) &=& \frac{9}{8\pi} g_B^2 \, \cos^2 \theta_B \, \frac{M_\Psi^2}{M_{Z_B}^2} M_{h_B} \left( 1 - 4 \frac{M_\Psi^2}{M_{h_B}^2}\right)^{3/2},\\[1.2ex]
\Gamma(h_B \to \bar \Sigma^+ \Sigma^+) &=&  \frac{9}{8\pi} g_B^2 \, \cos^2 \theta_B \frac{M_{\Sigma^+}^2}{M_{Z_B}^2} M_{h_B} \left( 1 - 4 \frac{M_{\Sigma^+}^2}{M_{h_B}^2}\right)^{3/2},\\[1.2ex]
\Gamma (h_B \to  \Sigma^0 \Sigma^0 ) &=& \frac{9}{16\pi}g_B^2 \, \cos^2 \theta_B \frac{M_{\Sigma^0}^2}{M_{Z_B}^2} M_{h_B} \left( 1 - 4 \frac{M_{\Sigma^0}^2}{M_{h_B}^2}\right)^{3/2}, \\[1.2ex]
\Gamma (h_B \to  \chi \chi ) &=& \frac{9}{16 \pi} g_B^2 \,  \cos^2 \theta_B \, \frac{M_{\chi}^2}{M_{Z_B}^2} M_{h_B} \left( 1- 4 \frac{M_\chi^2}{M_{h_B}^2}\right)^{3/2}, \\[1.2ex]
\Gamma (h_B \to h\,h ) &=& \frac{\sqrt{M_{h_B}^2-4M_h^2}}{32 \pi M_{h_B}^2} |c_{112}|^2 .
\end{eqnarray*}
The couplings between the gauge bosons and the new anomalons are given by (see Feynman rules in Appendix~\ref{sec:appFR}),
\begin{equation*}
 g_{W\Sigma} = - \frac{e }{\sin \theta_\text{W}} = - \sqrt{2} \,  g_{W\Psi}, \quad g_{Z\Psi_0} = -\frac{e}{\sin 2\theta_W}, \quad g_{Z\Psi^+} = \frac{e}{\tan 2 \theta_W}, \quad g_{Z\Sigma^+} = \frac{e}{\tan \theta_W},
 \end{equation*}
 whereas the trilinear interaction of Higgses in the scalar potential is parametrized by
 \begin{eqnarray*}
 c_{112} &=& -6 \lambda_h v_0 \cos^2 \theta_B \sin \theta_B + 6 \lambda_B v_B \cos \theta_B \sin^2 \theta_B \\[1ex]
 & & + \lambda_{HB} ( v_B \cos^3 \theta_B + 2 v_0 \cos^2 \theta_B \sin \theta_B -2v_B \cos \theta_B \sin^2 \theta_B - v_0 \sin^3 \theta_B ),
 \end{eqnarray*}
and the couplings from the scalar potential can be expressed as a function of the Higgs masses, vacuum expectation values and the mixing angle as follows:
 \begin{eqnarray}
\lambda_H &=& \frac{1}{2 v_0^2}(M_h^2 \cos^2\theta_B + M_{h_B}^2 \sin^2\theta_B),\\
\lambda_B &=& \frac{1}{2 v_B^2}(M_h^2 \sin^2 \theta_B + M_{h_B}^2 \cos^2\theta_B), \\
\lambda_{HB} &=& \frac{1}{v_0 v_B}(M_{h_B}^2-M_{h}^2)\sin \theta_B \cos \theta_B.
\end{eqnarray}

\bibliographystyle{JHEP}
\bibliography{Higgs-U1B}{}
\end{document}